\documentclass{aa}  

\makeatletter
\renewcommand*\aa@pageof{, page \thepage{} of \pageref*{LastPage}}
\makeatother

\usepackage{graphicx}
\usepackage{xcolor}
\usepackage{amsmath}
\usepackage{txfonts}
\usepackage[colorlinks, citecolor=blue, linkcolor=blue, urlcolor=blue]{hyperref}
\usepackage{multicol}
\usepackage{enumitem}
\usepackage[section]{placeins}

\newcommand*\Teff{$T_{\rm eff}$}
\newcommand*\logg{$\log{g}$}
\newcommand*\Vt{$v_{\rm t}$}
\newcommand*\gaia{\textit{Gaia~}}

\begin{document}

   \title{Atmospheric parameters of Cepheids from flux ratios with ATHOS\thanks{Based on observations collected at the European Southern Observatory under ESO programmes 66.D-0571(A), 072.D-0419(A), 072.D-0419(B), 072.D-0419(C), 073.D-0136(A), 073.D-0136(B), 074.D-0008(B), 076.B-0055(A), 081.D-0928(A), 082.D-0792(A), 082.D-0901(A), 089.D-0767(C), 091.D-0469(A), 097.D-0150(A), 098.D-0379(A), 099.D-0380(A), 0100.D-0273(A), 0100.D-0339(B), 0100.D-0397(A), 0101.D-0551(A), 0101.D-0697(A), 190.D-0237(A), 190.D-0237(E), 190.D-0237(F), and 266.D-5655(A).}}

   \subtitle{I. The temperature scale}

   \author{Bertrand Lemasle\inst{1}\fnmsep\thanks{Equal first authors}
          \and
          Michael Hanke\inst{1}\fnmsep$^{\star\star}$
          \and
          Jesper Storm\inst{2}
          \and
          Giuseppe Bono\inst{3,4}
          \and
          Eva K. Grebel\inst{1}}

   \institute{Astronomisches Rechen-Institut, Zentrum für Astronomie der Universität Heidelberg, Mönchhofstr. 12-14, D-69120 Heidelberg, Germany\\
              emails: \href{mailto:lemasle@uni-heidelberg.de}{lemasle@uni-heidelberg.de},
              \href{mailto:mhanke@ari.uni-heidelberg.de}{mhanke@ari.uni-heidelberg.de}.
         \and
          Leibniz-Institut für Astrophysik Potsdam (AIP), An der Sternwarte 16, D-14482 Potsdam, Germany
         \and Department of Physics, Università di Roma Tor Vergata, via della Ricerca Scientifica 1, I-00133 Roma, Italy
         \and INAF-Osservatorio Astronomico di Roma, via Frascati 33, I-00040 Monte Porzio Catone, Italy
          }

   \date{Received April 28, 2020; accepted June 25, 2020}

 
  \abstract
   {The effective temperature is a key parameter governing the properties of a star. For stellar chemistry, it has the strongest impact on the accuracy of the abundances derived. Since Cepheids are pulsating stars, determining their effective temperature is more complicated that in the case of non-variable stars.}
   {We want to provide a new temperature scale for classical Cepheids, with a high precision and full control of the systematics.}
   {Using a data-driven machine learning technique employing observed spectra, and taking great care to accurately phase single-epoch observations, we have tied flux ratios to (label) temperatures derived using the infrared surface brightness method.}
   {We identified 143 flux ratios that allow us to determine the effective temperature with a precision of a few~K and an accuracy better than 150~K, which is in line with the most accurate temperature measures available to date. The method does not require a normalization of the input spectra and provides homogeneous temperatures for low- and high-resolution spectra, even at the lowest signal-to-noise ratios. Due to the lack of a dataset of sufficient sample size for Small Magellanic Cloud Cepheids, the temperature scale does not extend to Cepheids with [Fe/H]~$<-$0.6~dex but nevertheless provides an exquisite, homogeneous means of characterizing Galactic and Large Magellanic Cloud Cepheids.}
   {The temperature scale will be extremely useful in the context of spectroscopic surveys for Milky Way archaeology with the WEAVE and 4MOST spectrographs. It paves the way for highly accurate and precise metallicity estimates, which will allow us to assess the possible metallicity dependence of Cepheids' period-luminosity relations and, in turn, to improve our measurement of the Hubble constant H$_{0}$.
   }
   \keywords{Stars: variables: Cepheids -- stars: fundamental parameters -- techniques: spectroscopic -- Methods: data analysis}

   \maketitle
%
\section{Introduction}

\par The effective temperature, \Teff{}, is one of the basic quantities characterizing a star and, in particular, one of the fundamental parameters shaping its spectrum. It is crucial for the accurate determination of chemical abundances because a biased \Teff{} has a strong impact on the abundances derived, much larger than the surface gravity (\logg), for instance. Temperatures are often derived from empirical relations employing photometry, by tying colors to temperature scales \citep[e.g.,][]{Bell1989,Alonso1999} previously defined using for example the infrared flux method \citep[IRFM,][]{Blackwell1977}. In this technique, the combination of the integrated flux and the infrared flux in a given band enables one to derive the effective temperature and the angular radius of the star. 

\par However, such methods are also sensitive to other parameters (e.g., \logg) and can be strongly affected by unusual chemical compositions \citep[][and references therein]{Huang2015}. Moreover, they require accurate determinations of the interstellar extinction, which is always more difficult in the case of disk stars and especially for Cepheids since they are pulsating and located close to the plane of the Milky Way.\\ 

\par As far as Cepheids are concerned, many studies have tackled the \Teff{} issue. For instance, \cite{Pel1978} did so using Walraven photometry and \cite{Kurucz1975} models, \cite{Fernley1989b} using the IRFM method and \cite{Kurucz1979} models, and \cite{Kiss1998b} using Strömgren photometry and Kurucz models. \cite{Welch1994}, \cite{Fouque1997}, and \cite{Kervella2004c} calibrated optical and near-infrared surface-brightness (IRSB) relations anchored to interferometric measurements of the radius of Cepheids. \cite{Kervella2004g} report a remarkable agreement between the previous two Cepheid studies as well as with their own calibration of the surface-brightness relation for dwarf stars \citep{Kervella2004f}. Period-luminosity relations can then be determined using distances derived from IRSB relations \citep{Storm2011a,Storm2011b}. The Spectro-Photo-Interferometry of Pulsating Stars code \citep[SPIPS,][]{Merand2015} now combines direct interferometric measurements of the Cepheid angular diameter (and its variations along the pulsation phase) with all other observables available into the same Cepheid model \citep{Breitfelder2016}. The method has improved the accuracy and precision on several model parameters.\\

\par Below 8000~K, the effective temperature dependence of the hydrogen lines makes them excellent \Teff{} indicators, especially given their independence on \logg. In particular, the wings of the Balmer lines provide one of the most reliable \Teff{} indicators for FGK stars \citep{Fuhrmann1993, Hanke2018, Amarsi2018a}. However, in Cepheids, their distorted profiles \citep[e.g.,][]{Nardetto2008c} rules them out so far, in the absence of appropriate atmosphere models for pulsating stars.

\par The excitation equilibrium of \ion{Fe}{i} and \ion{Fe}{ii} lines is commonly used to derive the effective temperature of FGK stars, including in our own analyses of Cepheids \citep[e.g.,][]{Lemasle2007,Lemasle2008,Lemasle2017}, under the assumptions of hydrostatic and local thermodynamic equilibrium (LTE).
The first hypothesis obviously does not hold for pulsating variables and biases the results as recently verified for Cepheids by \cite{Vasilyev2017,Vasilyev2018}. The LTE hypothesis does not hold either \citep[e.g.,][]{Fuhrmann1998,Thevenin1999} and this is also true for Cepheids \citep[e.g.,][]{Vasilyev2019}.  Alternatively, line depth ratios (LDRs) of pairs of lines with different sensitivity to \Teff{} have been put forward as temperature indicators by, e.g., \cite{Gray1991,Gray1994}. The number of ratios specifically calibrated to derive \Teff{} of Cepheids has increased from 32 ratios in \cite{Kovtyukh2000} to 131 in \cite{Kovtyukh2007} and 257 in \cite{Proxauf2018}. LDRs pose highly precise observables that are capable of reproducing the relative temperature variations throughout the pulsation phase. Yet, an accurate absolute calibration of the scale is still pending \citep[e.g.,][]{Kovtyukh2000,Mancino2020} Employing spectroscopic \Teff{} indicators bears the major advantage that they are free of influences from reddening, an important asset in the case of classical Cepheids which are located close to the Galactic plane.\\  

\defcitealias{Hanke2018}{H18}
\par In this paper, we take advantage of the large sample of high-quality spectra that are available for Cepheids to establish a data-driven approach using flux ratios (FRs) as means to predict \Teff{}, in a similar approach as developed for stable stars by \citet[][henceforth H18]{Hanke2018} for the new code ATHOS (A Tool for HOmogenizing Stellar parameters). Sect.~\ref{data} describes the data preparation, in particular how labels, the meaningful information (here, \Teff) attached to each spectrum, were derived (Sect.~\ref{IRSB}). The method for selecting FRs (from the bluest wavelengths to the calcium triplet (CaT) region) is outlined in Sect.~\ref{method}. We report our results in Sect.~\ref{results} and discuss them in Sect.~\ref{discussion}.

\section{Data preparation}
\label{data}

\subsection{Stellar parameters labels from the IRSB technique}
\label{IRSB}

\par The near-infrared ($J$,$H$,$K$) surface-brightness (IRSB) method \citep{Fouque1997} is a Baade-Wesselink-type method that uses the pulsational variations in
radial velocity, color, and luminosity to determine precise distances and radii
of pulsating stars. It is based on the geometrical relation between the
distance, $d$, to the star, the radius, $R$, and angular diameter,
$\theta$ of the star at different pulsation phases, $\phi$, namely:
\begin{equation}
\theta(\phi) = 2R(\phi)/d = 2(R_0 + \Delta R(\phi))/d
\end{equation}
With many data points at different phases this linear relation can be
trivially solved for the stellar radius and distance.\\

The radius at a given phase is obtained by integrating the pulsational
velocity curve for the star where the pulsational velocity is determined
from the observed radial velocity by applying a suitable projection
factor, $p$. The $p$-factor is largely a geometrical factor but depends
also on limb darkening, the exact method used to derive the radial velocity \citep[see][]{Borgniet2019}, and other technicalities. \cite{Storm2011a} calibrated the $p$-factor empirically and provide a detailed discussion and references on this issue.

The surface-brightness in the $V$-band, $F_V$, is defined as: 
\begin{equation} F_V(\phi)
= 4.2207 - 0.1V_0(\phi) - 0.5\log \theta(\phi) 
\end{equation}
where $V_0$ is the de-reddened visual magnitude. It was shown by \citet{Fouque1997} that the surface-brightness is
well described by a linear relation in the near-infrared $(V-K)_{0}$ color.
\cite{Storm2011a} adopted the calibration by \cite{Kervella2004g}:
\begin{equation} 
F_V = -0.1336(V-K)_0 + 3.9530 
\end{equation}
This calibration is based on interferometric angular diameters of Cepheids.\\

\par To determine \Teff($\phi$) from the Stefan-Boltzmann equation,
in addition to the stellar radii just determined, it is necessary
to determine the luminosity at each phase point given the absolute $V$-band magnitude $M_V$, and the
bolometric correction. $M_V$ is determined from the $V$-band light curve given the distance from above and the reddening. The bolometric corrections have been computed from the 2012 revised (ODFNEW\footnote{\url{http://wwwuser.oats.inaf.it/castelli/colors/bcp.html}}) 
ATLAS9 fluxes of \citet{Bessell1998} based on \citet{Castelli2003} models, given the metallicity, \logg, and \Teff. As \Teff{} is needed to enter the table it was necessary to make a few (usually only one) iterations to
make \Teff{} converge.

\par The surface gravity \logg{} depends on radius, the adopted mass for the Cepheid, and at
any given phase also the additional acceleration from the dynamical
atmosphere.  A mass of 6$\vec{M}_\odot$ was adopted in 
\cite{Storm2011a} for all the stars.  The exact choice does not make a significant difference on the derived temperature. Had a mass of 8$\vec{M}_\odot$ been adopted then the value of \Teff{} would have changed by less than a degree.

The adopted metallicity also affects the bolometric correction and thus
the resulting \Teff{} but the effect is very weak. An error in
the adopted metallicity of 0.5~dex would lead to an error in \Teff{} of only 15~K.

Similarly, the adopted $p$-factor for the conversion between the observed
radial velocity and pulsation velocity has no effect on the derived
temperature as the resulting change in radius is exactly compensated by
the changed distance and thus in the luminosity.\\

As shown already by \citet{Fouque1997}, the distance and radius estimates are
very robust to errors in the adopted reddenings. On the other hand, $M_V$
depends directly on the reddening so the estimated temperatures are
indeed sensitive to the adopted reddening.  The effect of artificially
lowering the adopted reddening, $E(B-V)$, by 0.05~mag is to reduce the
estimated temperature by 150 to 200~K depending on the pulsation period
of the Cepheid. This is the only significant systematic uncertainty
apart from the adopted surface-brightness relation.

\par By construction (see Sect.~\ref{train}), our spectroscopic dataset mostly consists of nearby Cepheids for which a wealth of external data is available. In particular, the values of the reddening, E(B-V), are as low as 0.003~mag for $\zeta$~Gem and below 0.2~mag for 46\% of our sample.
This ensures that the values of the reddening are either negligible or well-constrained in order to limit systematic errors on the labels. Our original sample also includes Cepheids in the Large Magellanic Cloud (LMC) and in the Small Magellanic Cloud (SMC) for which the E(B-V) values were determined using the reddening maps of \cite{Gorski2020}. E(B-V) values range from 0.081 to 0.175~mag for our LMC Cepheids and from 0.058 to 0.103~mag for our SMC Cepheids, with typical uncertainties of 0.005~mag.
    
\par The reddening value adopted is crucial for the determination of \Teff{} via the IRSB method. A simple test for the 7.6~day period Cepheid W~Sgr showed that the estimated temperature is reduced by 190~K when $E(B-V)$ is artificially lowered by 0.05~mag with respect to the adopted value of 0.108~mag, whereas for the 41~day Cepheid RS~Pup such a change of the reddening reduces the estimated temperature by 150~K. Table~\ref{sample} indicates that the reddening values adopted in our IRSB analysis (used in \cite{Storm2011a} and based on the \cite{Fouque2007} compilation) match closely those tabulated by \cite{Laney2007}. They are on the photometric system of \cite{Laney2007} as described in \cite{Fouque2007}. \cite{Laney2007} showed that externally determined reddenings (for instance, using main-sequence stars that either are companions of the Cepheid or belong to the same open cluster) fall on the same scale as those derived from multicolor photometry, hence providing a homogeneous reddening scale.

\par \cite{Fouque2007} attempted to combine Cepheids' reddenings obtained from the 1970's on by different groups onto the system of \cite{Laney2007} using linear transformations. From their Table~1, the scatter around the different transformations reaches 0.04~mag, while for the largest sample in that study \citep{Fernie1995b}, it is only 0.03~mag. From the same table, it appears that the Dean Warren and Cousins system \citep{Dean1978} is quite close to the \cite{Laney2007} system, while most of the other studies\footnote{The study by \cite{Eggen1996} is an outlier with a slope significantly different from unity} seem to cluster around similar values with a systematic offset of $\sim$0.04~mag. Assuming two types of systems, one could claim systematic uncertainties of half of the offset or 0.02~mag. This is probably a too optimistic view and we adopt instead a more conservative value of 0.03~mag. It is worth mentioning that \cite{Turner2016} using space reddenings reported that the values from \citet{Storm2011a} should be transformed linearly with an offset of +0.05~mag for a typical $E(B-V)=0.5$~mag.
\par Numerous studies have focused on the reddening scale in the Magellanic Clouds. We used the maps obtained by \citet{Gorski2020}, that have a systematic uncertainty of the average reddening value of 0.013~mag. A comparison with previous studies in the LMC, for instance, indicates a mean difference of 0.061~mag, with a standard deviation of 0.012~mag between the reddening values from \cite{Gorski2020} and those from \cite{Haschke2011}. The same comparison between the results of \citet{Gorski2020} and those of \cite{Inno2016} leads to an average offset of 0.004~mag with a standard deviation of 0.087~mag.
\par Reddenings are by far the main source of systematic uncertainties in our study, and improving the accuracy of Cepheids' reddenings will provide an exquisite accuracy for our flux ratio method.

\subsection{Spectroscopic data}
\label{spectra}

\par For training and testing the method, we collected a large number of high-resolution spectra at high signal-to-noise ratios (S/N). We limited ourselves to a small number of spectrographs, namely HARPS \citep{Mayor2003} at the La Silla 3.6m telescope, HARPS-N \citep{Cosentino2012} at the Telescopio Nazionale Galileo (TNG), UVES \citep{Dekker2000} at the Very Large Telescope (VLT), FEROS \citep{Kaufer1999} at the La Silla 2.2m telescope, and STELLA \citep{Strassmeier2004,Weber2012} at the 1.2m telescope at Izana Observatory. We considered only spectra with S/N in excess of 50~pixel$^{-1}$ for Galactic Cepheids, but had to weaken this requirement to a S/N of 30~pixel$^{-1}$ for Magellanic Cepheids in order to be able to incorporate them in our sample\footnote{The S/N for HARPS spectra is either well above (nearby stars), or well below (spectra dedicated to radial velocity determination) this threshold.}. Table~\ref{spectrographs} lists the properties of the spectrographs and the number of spectra gathered for each of them. Table~\ref{sample} lists the Cepheids in the sample and the number of spectra gathered for each instrument.

\begin{table}[ht]
    \caption{Characteristics of the spectra used in this study.}
    \label{spectrographs}
    \centering
    \resizebox{\columnwidth}{!}{%
    \begin{tabular}{lcrc}
    \hline\hline
    &&&\\[-5pt]
        Spectrograph & Wavelength & \multicolumn{1}{c}{Resolving} & Number of \\
                     & coverage  & \multicolumn{1}{c}{power} & spectra \\
                     & [\AA] & & \\[3pt]
    \hline
    &&&\\[-5pt]
         FEROS   & 3530-9220 & 48\,000  &   1\\
         HARPS   & 3780-6910 & 115\,000 & 274\\
         HARPS-N & 3780-6910 & 115\,000 & 104\\
         STELLA  & 3870-8700 & 55\,000  & 123\\
         UVES$^{(a)}$ & 3730-9464 & >34\,500 & 822\\[3pt]
    \hline
    \end{tabular}}
    \tablefoot{\tablefoottext{a}{For UVES, we employed spectra from a variety of setups with different dichroics. Hence, we only show the extrema in terms of wavelength coverage and the minimum resolving power. Note that UVES dichroic observations are counted here as one spectrum, while the blue and red arm spectra appear in two different files in the ESO archive.}}
\end{table}

\par Spectra that were taken with FEROS, HARPS, and UVES were retrieved from the ESO (European Southern Observatory) archive. The spectra taken with HARPS-N were taken from the TNG archive. HARPS and HARPS-N products were used in the 1d spectral data format where individual spectral orders are merged. The STELLA spectra were collected by J. Storm. Since co-addition of the spectral orders is not included in the STELLA pipeline, we performed this operation using a S/N-weighted sum in the spectral regions where orders overlap.\\

\par One of the strengths of the our FR-based approach is that it does not rely on normalized spectra by restricting itself to small wavelength ranges for which the assumption of constant continua holds. For preparing the data, we thus only had to place the spectra in the stellar rest frame: we corrected the spectra for line-of-sight motions using a cross-correlation method with three different, synthetic template spectra that are representative of Cepheids and that were selected according to the temperature of the star at the time of the spectroscopic observation. The templates' characteristics are listed in Table~\ref{Vr_templates}. Finally, we performed a simple cosmic ray rejection by interpolating over features showing strong variances on scales much smaller than the width of the instrumental profile.   

\begin{table}[ht]
\centering

    \caption{Characteristics of the synthetic template spectra used for deriving the radial velocity using a cross-correlation method.}
    \label{Vr_templates}
    \begin{tabular}{rcccc}
     \hline \hline
&&&&\\[-5pt]
        \multicolumn{1}{c}{\Teff{} range} & \Teff & \logg & [Fe/H] & \Vt \\
                      &  [K]  & [dex] &  [dex] &[km\,s$^{-1}$]\\[3pt]
    \hline
&&&&\\[-5pt]
           \Teff{} $<$ 5250~K & 4800 & 0.50 & +0.1 & 2.0 \\ 
5250~K $<$ \Teff{} $<$ 6000~K & 5800 & 1.00 & +0.1 & 2.0 \\
           \Teff{} $>$ 6000~K & 6600 & 1.30 & +0.1 & 2.0 \\[3pt]
    \hline    
    \end{tabular}
\end{table}

\begin{table*}[ht]
\footnotesize{
\centering
    \begin{tabular}{rlccrrrrr}
    \hline \hline
&&&&\\[-5pt]
        \multicolumn{1}{c}{Star} & \multicolumn{1}{c}{$\log{P}$} & $E(B-V)_{\text{IRSB}}$ & $E(B-V)_{\text{L07}}$ &\multicolumn{5}{c}{Spectrograph}\\
             &       & [mag]  & [mag]  &  \multicolumn{1}{c}{F}  &  \multicolumn{1}{c}{H}  &  \multicolumn{1}{c}{N}  &  \multicolumn{1}{c}{S}  &  \multicolumn{1}{c}{U}  \\ [3pt] 
    \hline
&&&&&&\\[-5pt]
      RT Aur & 0.571489 & 0.059 & 0.079 & $\cdots$ & 3 & $\cdots$ & 3 & 89 \\
      AD Gem & 0.578408 & 0.173 & 0.382 & $\cdots$ & $\cdots$ & $\cdots$ & $\cdots$ & 1 \\
      BF Oph & 0.609329 & 0.235 & 0.223 & $\cdots$ & 2 & $\cdots$ & 3 & 5 \\
       T Vul & 0.646934 & 0.060 & 0.052 & $\cdots$ & 3 & $\cdots$ & 26 & $\cdots$ \\
      FF Aql & 0.65039  & 0.196 & 0.188 & $\cdots$ & $\cdots$ & $\cdots$ & 27 & $\cdots$ \\
$\delta$ Cep & 0.729678 & 0.075 & 0.073 & $\cdots$ & $\cdots$ & 104 & 18 & $\cdots$ \\
      BB Sgr & 0.821971 & 0.281 & 0.302 & $\cdots$ & 2 & $\cdots$ & 3 & 5 \\
       U Sgr & 0.828997 & 0.402 & 0.408 & $\cdots$ & 2 & $\cdots$ & 3 & 60 \\
       W Sgr & 0.880529 & 0.108 & 0.450 & $\cdots$ & 3 & $\cdots$ & $\cdots$ & $\cdots$ \\
       S Sge & 0.923352 & 0.099 & 0.084 & $\cdots$ & $\cdots$ & $\cdots$ & 19 & $\cdots$ \\
 $\beta$ Dor & 0.993131 & 0.051 & 0.042 & 1 & 46 & $\cdots$ & $\cdots$ & 78 \\
 $\zeta$ Gem & 1.006497 & 0.014 & 0.003 & $\cdots$ & 47 & $\cdots$ & 3 & 85 \\
       Z Sct & 1.110645 & 0.492 & 0.450 & $\cdots$ & $\cdots$ & $\cdots$ & $\cdots$ & 9 \\      
      TT Aql & 1.138459 & 0.435 & 0.417 & $\cdots$ & 2 & $\cdots$ & 3 & 50 \\
       Y Oph & 1.233609 & 0.647 & 0.450 & $\cdots$ & 7 & $\cdots$ & $\cdots$ & $\cdots$ \\
      RU Sct & 1.29448  & 0.911 & 0.924 & $\cdots$ & $\cdots$ & $\cdots$ & 2 & 43 \\
      RY Sco & 1.307927 & 0.718 & 0.746 & $\cdots$ & 3 & $\cdots$ & 3 & 9 \\
      RZ Vel & 1.309564 & 0.300 & 0.294 & $\cdots$ & 7 & $\cdots$ & $\cdots$ & 48 \\
      WZ Sgr & 1.339443 & 0.435 & 0.450 & $\cdots$ & 1 & $\cdots$ & 3 & 9 \\
      VZ Pup & 1.364945 & 0.455 & 0.450 & $\cdots$ & $\cdots$ & $\cdots$ & 1 & 60 \\
       T Mon & 1.431915 & 0.179 & 0.165 & $\cdots$ & 10 & $\cdots$ & 3 & 55 \\
     $l$ Car & 1.550816 & 0.146 & 0.138 & $\cdots$ & 111 & $\cdots$ & $\cdots$ & 80 \\
       U Car & 1.58897  & 0.263 & 0.260 & $\cdots$ & 6 & $\cdots$ & $\cdots$ & 50 \\
      RS Pup & 1.61742  & 0.457 & 0.454 & $\cdots$ & 19 & $\cdots$ & 3 & 54 \\[3pt]
    \hline
&&&&&&\\[-5pt]    
      HV 6093 & 0.679881 & 0.081 & $\cdots$ & $\cdots$ & $\cdots$ & $\cdots$ & $\cdots$ & 4 \\ 
      HV 2405 & 0.840333 & 0.114 & $\cdots$ & $\cdots$ & $\cdots$ & $\cdots$ & $\cdots$ & 2 \\ 
     HV 12452 & 0.941454 & 0.175 & $\cdots$ & $\cdots$ & $\cdots$ & $\cdots$ & $\cdots$ & 2 \\ 
{\it HV 1335} & 1.157800 & 0.093 & $\cdots$ & $\cdots$ & $\cdots$ & $\cdots$ & $\cdots$ & 3 \\ 
{\it HV 1328} & 1.199692 & 0.058 & $\cdots$ & $\cdots$ & $\cdots$ & $\cdots$ & $\cdots$ & 3 \\ 
{\it HV 1333} & 1.212084 & 0.100 & $\cdots$ & $\cdots$ & $\cdots$ & $\cdots$ & $\cdots$ & 3 \\ 
 {\it HV 822} & 1.223807 & 0.086 & $\cdots$ & $\cdots$ & $\cdots$ & $\cdots$ & $\cdots$ & 3 \\ 
      HV 1023 & 1.424235 & 0.121 & $\cdots$ & $\cdots$ & $\cdots$ & $\cdots$ & $\cdots$ & 2 \\ 
       HV 879 & 1.566167 & 0.146 & $\cdots$ & $\cdots$ & $\cdots$ & $\cdots$ & $\cdots$ & 2 \\ 
 {\it HV 837} & 1.630509 & 0.076 & $\cdots$ & $\cdots$ & $\cdots$ & $\cdots$ & $\cdots$ & 2 \\ 
       HV 877 & 1.655215 & 0.128 & $\cdots$ & $\cdots$ & $\cdots$ & $\cdots$ & $\cdots$ & 2 \\ 
      HV 2369 & 1.684646 & 0.142 & $\cdots$ & $\cdots$ & $\cdots$ & $\cdots$ & $\cdots$ & 2 \\ 
      HV 2827 & 1.896354 & 0.12  & $\cdots$ & $\cdots$ & $\cdots$ & $\cdots$ & $\cdots$ & 2 \\[3pt] 
    \hline
    \end{tabular}
    \caption{Properties of the Cepheids in the dataset: logarithm of the pulsation period ($P$), $E(B-V)$ from \cite{Storm2011a} used for the IRSB analysis, $E(B-V)$ tabulated by \cite{Laney2007}. The table also details the spectroscopic data available for these stars (F: FEROS, H: HARPS, N: HARPS-N, S: STELLA, U:UVES). The upper panel lists Galactic Cepheids, the lower panel Magellanic Cepheids. Note that SMC stars (identified in italics) have not been used for training or testing, but only for assessing that the current relations do not hold anymore at the metallicities of SMC Cepheids (see Sect.~\ref{low_met}).}
    \label{sample}
}
\end{table*}

\subsection{Training at a resolving power of $R=5000$:\\ pros and cons}
\label{Train_res}

\par All the spectra were then rebinned to a common resolving power of $R=5000$. Training at such a low resolution limits the number of available relations when compared to training at higher resolution ($R>>10\,000$). This is not a problem since the accuracy remains driven by the intrinsic accuracy of the labels, and the precision remains better than any other method available.

\par Training at $R=5000$ comes instead with two indisputable advantages. Firstly, the already high S/N of our spectra reaches extremely high (>1000) values when down-sampled to $R\sim5000$, which makes the training insensitive to photon shot noise (this also applies to unseen data, since high-resolution spectra (outside of the training sample) would need to be degraded to the resolving power $R=5000$ in order to derive \Teff{} using the method developed in this study). Secondly, in the perspective of large spectroscopic surveys with, e.g., WEAVE\footnote{William Herschel Telescope Enhanced Area Velocity Explorer} \citep[][]{Dalton2016}, 4MOST\footnote{4-metre Multi-Object Spectrograph Telescope
} \citep[][]{deJong2019}, or MOSAIC\footnote{Multi-Object Spectrograph for Astrophysics, Intergalactic-medium studies and Cosmology} \citep[][]{CJEvans2015}, it provides the same temperature scale for samples studied at low (of the order of $R=5000$) or medium resolving power (of the order of $R=20\,000$).\\
\par However, training at lower resolving power further comes with the caveat of being in all generality more metallicity sensitive because of blending features. This is not problematic within the range of parameters covered by the training sample, as it is intrinsically accounted for, but limits the applicability of the method to this exact same range (which is anyway always the case when using machine-learning techniques). In case of varying [X/Fe] among targets, this very circumstance poses an additional potential bias. It should be noted that in Cepheids, despite star-to-star variations, \cite{Genovali2015} reported that [$\alpha$/Fe] shows a flat distribution across the entire Galactic thin disk (as one would expect for a young population), and no trend with $\log{P}$. As far as neutron-capture elements in Cepheids are concerned, \cite{daSilva2016} found a positive slope for [La/Fe], [Ce/Fe], [Nd/Fe], and [Eu/Fe]  but no slope for [Y/Fe]. The same authors also reported no trend (except maybe for Ce) as a function of $\log{P}$. Including a large number of Magellanic Cepheids in the sample is a strong requirement for improving the universality of the method.

\par At the beginning of Sect. \ref{Train_res}, we have exposed our reasons for training at $R=5000$. Determining \Teff{} for an unseen Cepheid spectrum at any resolving power R>5000 then requires that this spectrum is first degraded to the training resolving power (as mentioned above, no continuum normalization is required). It goes without saying that the method can therefore be applied only to spectra with $R\geq5000$. By construction, it does not loose generality with varying resolution or \Teff{}. This could instead be the case for LDRs \citep[see][]{Mancino2020} because some line could disappear or become blended at a given \Teff{} which may itself depend on the resolution of the spectrum.

\par We set a higher priority on our main goal (providing a homogeneous \Teff{} scale for low- and high-resolution Cepheid spectra in upcoming surveys) than on potentially optimizing accuracy and precision via a different choice for the training resolution. Moreover, in the perspective of a canonical chemical analysis, we are fully satisfied with the accuracy and precision reached, as will be detailed in Sect.~\ref{results}. A detailed analysis of the performances of flux ratios as a function of resolution will therefore not be undertaken in this paper. We anticipate that training at higher resolution will increase the number of calibrating relations and therefore the precision.

\subsection{The need for an accurate phasing of the spectra}
\label{phasing}

\par
\begin{figure}
    \centering
    \resizebox{\hsize}{!}{\includegraphics{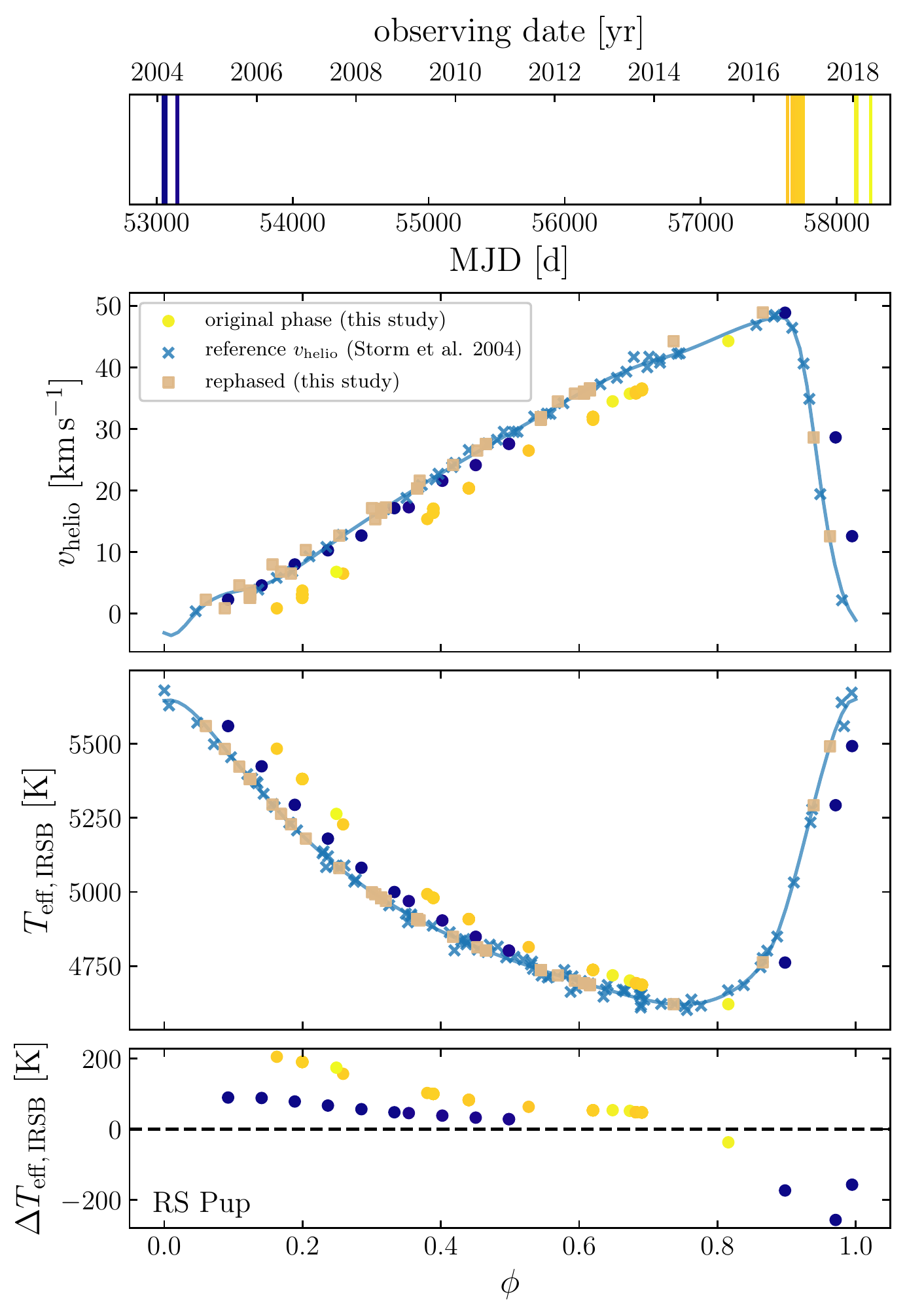}}
      \caption{Impact of erroneous phasing of individual spectra on the \Teff{} labels for the Cepheid RS~Pup: the \textit{top panel} displays the MJD of the observation for individual spectra. The \textit{second panel} shows our individual radial velocity measurements (circles that are color-coded by MJD, cf., \textit{upper panel}) versus the radial velocity data used for the IRSB analysis \citep[blue crosses,][]{Storm2004b}. Spectra rephased to match the radial velocity curve are shown as pale brown squares. Using the same color coding, the \textit{third} and \textit{fourth panels} display the \Teff{} variation, and the \Teff{} offset between original labels and their corrected value after rephasing, respectively.}
      \label{Fig:rephasing_demo_RSPup}
\end{figure}

 We quickly realized during the training process that even small errors in the phasing (of the order of a few hundredths of the period) of the training spectra could lead to relatively large biases (up to a few 100~K in the labels, see Fig.~\ref{Fig:rephasing_demo_RSPup}), and that these biases would be the parameter limiting the accuracy of our temperatures. Here we want to stress that the phase of the observations is only used to attribute \Teff{} labels to the spectra employing values derived via the IRSB method, and is not used further in the process.\\

\par The phase of the spectroscopic observations was initially determined using the Modified Julian Date (MJD) at which the spectra were taken, together with the exact same values of reference epoch and pulsation period that were used in the IRSB analysis. An extremely large (and constantly updated) collection of radial velocity measurements (spanning three decades) has been collected to apply the IRSB method \citep[see, for instance,][]{Storm2011a}. To ensure that our high-resolution spectra (spanning 15 years) are placed on the exact same phase scale as the IRSB analysis, we checked our own radial velocity measurements against the IRSB radial velocity curves and phase shifts were applied to ensure a consistent phasing between both datasets. It is no surprise that they concerned primarily:
\begin{itemize}[nosep]
    \item stars for which the high-resolution spectra were taken at very different epochs from the spectra used for the IRSB analysis; 
    \item Cepheids with known period changes. These period changes can be related to secular evolution of the Cepheids across the instability strip \citep[e.g.,][]{Berdnikov2019} or to cycle-to-cycle variations that affect, in particular, long-period Cepheids \citep{Anderson2014b,Anderson2016a,Anderson2016d}. 
\end{itemize}   
\par An alternative approach would have been to take into account period drifts. However, data are often too scarce to derive them accurately, and it is usually difficult to reach a consensus on the value of even a simple, linear period drift \citep[see][and references therein]{Breitfelder2016}. To our knowledge, RS~Pup is the only Cepheid for which the modulation of the period over several decades has been derived accurately, using a fifth-degree polynomial \citep{Kervella2017}. Given the current insufficient accuracy of period drifts, we adopted instead the method described above. We cannot exclude that a fraction of the remaining residuals discussed in Sect. \ref{res_teff} originates from a still inaccurate phasing of some spectra.\\

\par For a few stars, an accurate rephasing based on the radial velocity curve proved impossible. It is the case of the exceptional Cepheid X~Sgr, where the signature of strong shock waves (emission, line doubling) has been reported by \cite{Mathias2006}. In a few other stars (S~Mus, EV~Sct, U~Aql, V~Car, V~Cen, and V350~Sgr), a strong orbital motion was detected, which limited the accuracy of the phasing, and these stars were therefore not further considered despite a good number of spectra available. Finally, a few other spectra were rejected because the radial velocity curves of the stars were of lower quality (possibly due to orbital motion as well) and/or with incomplete phase coverage (KQ~Sco, RZ~Vel: spectra between phases 0.9 and 0.1, S~Nor, T~Vel, UZ~Sct, WZ~Car, XX~Sgr, Y~Sgr).

\subsection{The dataset}
\label{train}

\par A good dataset must have the following characteristics:
\begin{itemize}[nosep]
    \item the labels are accurate; 
    \item they cover the entire parameter space and are homogeneously distributed, not only in label space (\Teff{}), but also in other parameters that may affect an FR (e.g., [Fe/H] and \logg{}); 
    \item a large number of spectra is available.
\end{itemize}
However, as with many if not all data-driven methods, it is close to impossible to generate such a dataset. Here we investigate whether the conditions listed above are satisfied.\\

\par We retained 24 Galactic Cepheids for our sample. They are listed in Table~\ref{sample}, together with the number of spectra used for each of them. We considered only stars for which a good number ($>$10) of spectra were available and for which accurate phases could be determined (see Sect.~\ref{phasing}). We note in particular that a few stars are represented by a considerable number of spectra, namely $l$~Car (191), $\zeta$~Gem (135), $\beta$~Dor (125), $\delta$~Cep (122), and RT~Aur (95). We end up with a total of 1324 spectra.\\

\par Although in the Magellanic Clouds only few spectra per star are available, 32 spectra of 13 Magellanic Cepheids were also added to the dataset. Including a large number of spectra of LMC and especially SMC stars in the sample would present the advantage that metallicity-dependent relations would naturally be excluded\footnote{In practice, the same number of spectra for metal-rich (typically Galactic) and metal-poor (typically SMC) Cepheids would be required. Such a sample of SMC Cepheids does not currently exist. An alternative would be to attribute very high weights to the SMC spectra, which comes with the caveat that the labels need to be extremely well constrained and the spectra need to have a high S/N and further be free of artefacts such as cosmic-ray hits. Unfortunately, none of the latter is satisfied for the spectra at hand.}. However, for these stars we only have few spectra, which in addition do not cover the entire wavelength range. This means that temperatures derived from flux ratios located in different spectral domains would be differently affected by possible metallicity dependences. This does not seem to be a satisfying outcome. We decided instead to train at high metallicity only and to check for \Teff{} consistency of individual relations at low metallicity  (see Sect.~\ref{low_met}).\\

\par The first condition we set for our sample is met: we selected bright, nearby Cepheids that were thoroughly investigated over the years. The radial velocity curves and the light curves are well sampled. The spectra were phased with great care. As a result, accurate labels could be derived following the method described in Sect.~\ref{IRSB}. 

\par Our sample covers a broad range of periods ($3.7~\mathrm{days}\leq P \leq 78.8~\mathrm{days}$, see Table \ref{sample}), which ensures a good coverage of the \Teff{}-\logg{} space. Indeed, the coolest stars in our sample, $l$~Car and RS~Pup, have temperatures ranging from as low as 4578~K to 5282~K (respectively from 4621~K to 5560~K). The hottest Cepheid in our sample, RT~Aur,  has temperatures ranging from 5573~K up to 6592~K. For these stars, \logg{} varies between 0.41~dex and 1.30~dex ($l$~Car),  between 0.07~dex and 1.39~dex (RS~Pup), and between 2.16~dex and 2.50~dex (RT~Aur). The other Cepheids have intermediate values for \Teff{} and \logg{}.\\
\begin{figure}
    \centering
    \resizebox{\hsize}{!}{\includegraphics{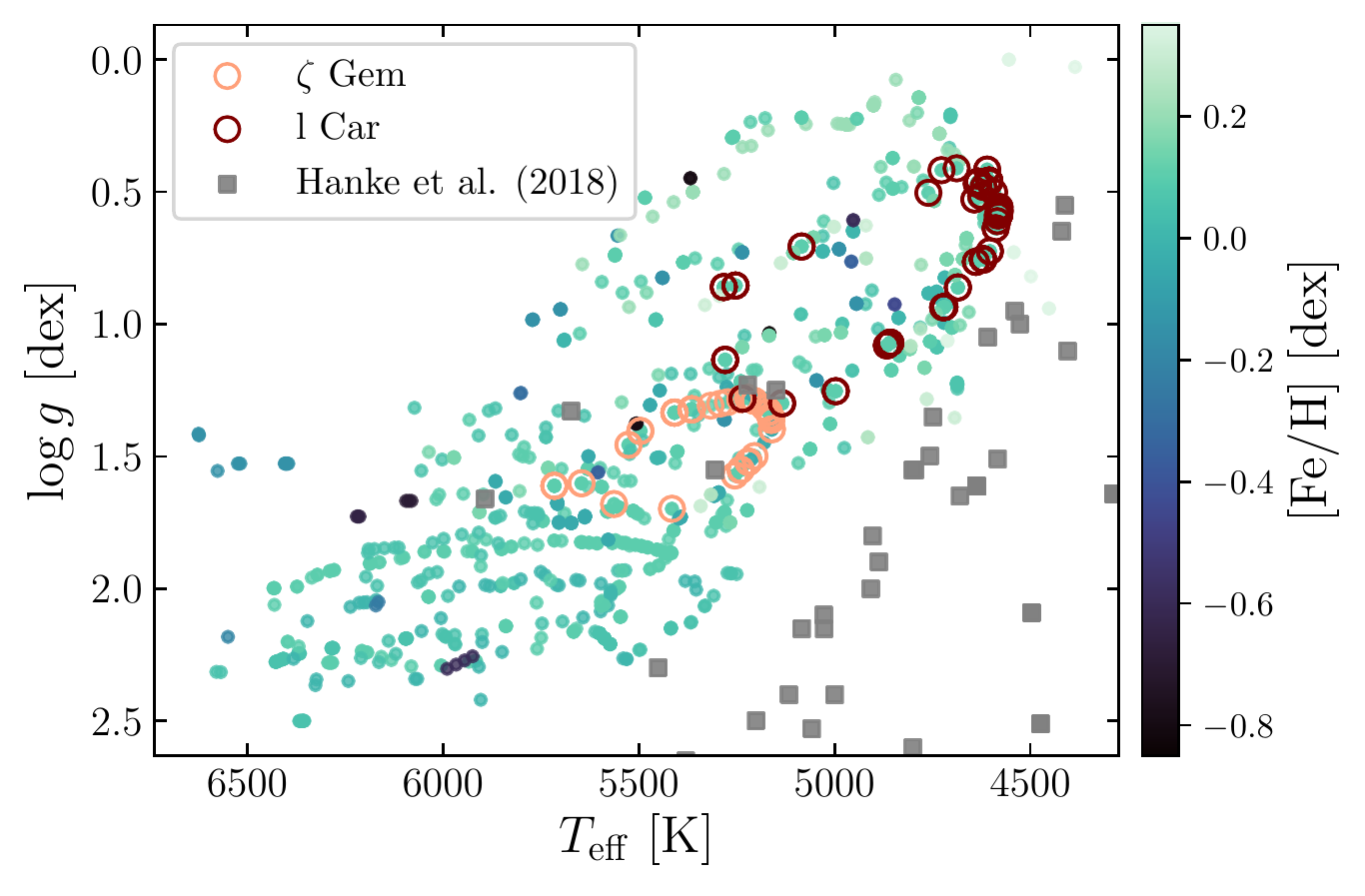}}
      \caption{Parameter space (\Teff, \logg, [Fe/H]) covered by the compiled sample of spectra. Highlighted by salmon and red circles are the Cepheids $\zeta$~Gem and  $l$~Car, for which we have the most spectra at hand (135 and 191, respectively). Gray squares indicate a small portion of the giants used in the ATHOS dataset for stable stars \citepalias{Hanke2018}.}
      \label{Fig:sample_HRD}
\end{figure}

\par However, the condition is not met as far as [Fe/H] is concerned. The best-studied stars are naturally among the closest Cepheids (which also allowed to collect large amounts of spectra in a reasonable amount of observing time). Given the presence of abundance gradients in the Milky Way \citep[e.g.,][]{Luck2011a,Yong2012,Genovali2013}, those Cepheids have metallicities close to Solar. Cepheids in the Milky Way span almost 1.0~dex in [Fe/H], from $\approx$+0.5~dex in the inner disk \citep[e.g.,][]{Pedicelli2010,Andrievsky2016} to $\approx$--0.5~dex in the outer disk \citep[e.g.,][]{Lemasle2008,Luck2011b,Genovali2014}. Due to their distances ($>$3~kpc) and their location close to the plane, less spectra are available for the stars located at the edges of the metallicity range, and the IRSB data is also scarce. To extend the metallicity coverage, a good alternative is to turn to the Magellanic Clouds, where a few tens of Cepheids have been studied spectroscopically at high resolution \citep{Romaniello2008,Lemasle2017} and for which IRSB analyses are available \citep{Storm2011b,Gieren2018}. [Fe/H] roughly ranges from --0.2 to --0.6~dex in the LMC and from --0.6 to --0.9~dex in the SMC. Unfortunately, very few medium- or high-resolution spectra are available, and even less when requiring S/N$>$30~pixel$^{-1}$. Moreover, all stars are represented by a unique spectrum taken at random phase, which means that we do not have at our disposal spectra covering a large range of \Teff, \logg. and periods for low-metallicity Cepheids.\\

\par The third condition is largely met: we have gathered a large number of spectra for each of the training Cepheids (see Table~\ref{sample}), namely a grand total of 1324 spectra, which provide a full phase coverage for each of the selected stars. The number of non-pulsating benchmark stars for covering the entire Hertzsprung-Russell diagram is considerably lower \cite[][\gaia benchmark stars]{Jofre2014}, \citet[][HD~20]{Hanke2020}.

\section{Method}
\label{method}

\par The approach pursued throughout this work closely follows the one described in \citetalias{Hanke2018}. To this extent, in order to enable the computationally efficient handling of the data, all spectra were binned to a common log-linear dispersion scale, where the $n$th element is expressed as
\begin{equation}
\log_{10}{\lambda}_n = \log_{10}{\lambda}_0 + n \cdot \delta \log_{10}{\lambda},\,\mathrm{for}\,n\in\{0,...,N\},    
\end{equation}
with $\lambda_0$ being the starting wavelength of 3800~{\AA}, $\delta \log_{10}{\lambda}$ the dispersion spacing, $2.7\cdot10^{-5}$, and $N$ the total number of pixels, here taken to be 13\,333. This way, our maximum considered wavelength is $\sim8705$~{\AA}. Employing a log-linear $\lambda$ scale bears the main advantage that profile widths remain approximately constant in pixel space\footnote{Amounting to $(5000\cdot(10^{\delta \log_{10}{\lambda}} - 1))^{-1}\approx3.22$~pixels from the instrumental resolution alone. We note that the effective line width may be larger owing to intrinsic line broadening mechanisms, such as rotation and/or macroturbulent motions in the stellar atmosphere.} over the extensive range used here.

\subsection{Feature selection}
\par We recall here the definition from \citetalias{Hanke2018} of a spectral FR between two wavelength windows that are centered at $\lambda_i$ and $\lambda_j$, respectively:
\begin{equation}
\label{Eq:FR}
 \mathrm{FR}_{\lambda_i,\lambda_j} = \frac{\langle{F}_{\lambda_i}\rangle}{\langle{F}_{\lambda_j}\rangle},
\end{equation}
where
\begin{equation}
\label{Eq:medflux}
\langle{F}_{\lambda_i} \rangle = \frac{1}{5}\sum_n F(\lambda_n), \quad n \in \{i - 2,...,i + 2\}
\end{equation}
is the mean flux of five pixels.\\ 

\par Each FR potentially qualifies as feature that encodes information about $T_\mathrm{eff}$, which in principle leaves us with $\mathcal{O}(10^{10})$ explanatory variables. This enormous number calls for an efficient and robust feature-selection process prior to the actual training and validation of feasible relations. Our first -- and arguably most important -- pre-selection criterion considered only those FRs for which the two flux windows are not separated by more than 30 pixels. While reducing the number of features by two orders of magnitude, this limit makes the FRs essentially scale-free, that is, normalization-independent, since it ensures that the continuum variation between the two closely-spaced regions of interest can be safely neglected (see Sect. \ref{cont var}). An additional applied criterion was that only those FRs were allowed as features that showed a maximum range of values in excess of 0.15. This was enforced to ensure that noise has a limited impact on the predicted labels from each FR (see also Sect. \ref{Sect: SNR stab}).\\

\par As a next step, we computed Pearson correlation coefficients, $\rho_{\mathrm{FR},T_\mathrm{eff}}$, between all remaining FRs and the \Teff{} labels associated to the underlying spectra. By requiring strong (anti-)correlations (i.e., $|\rho_{\mathrm{FR},T_\mathrm{eff}}|>0.93$) between the features and the labels, we further restricted the feature space to 33\,291 candidates. We emphasize at this point that by proceeding like this we potentially rejected a large number of tight feature-label relations, in particular those that are strongly non-linear and/or non-monotonic. This is owed to using computationally cheap correlation coefficients that test for linearity and could only be circumvented by investing substantially larger computational resources.\\

\subsection{Training and testing}
\begin{figure}[!h]
    \centering
    \resizebox{\hsize}{!}{\includegraphics{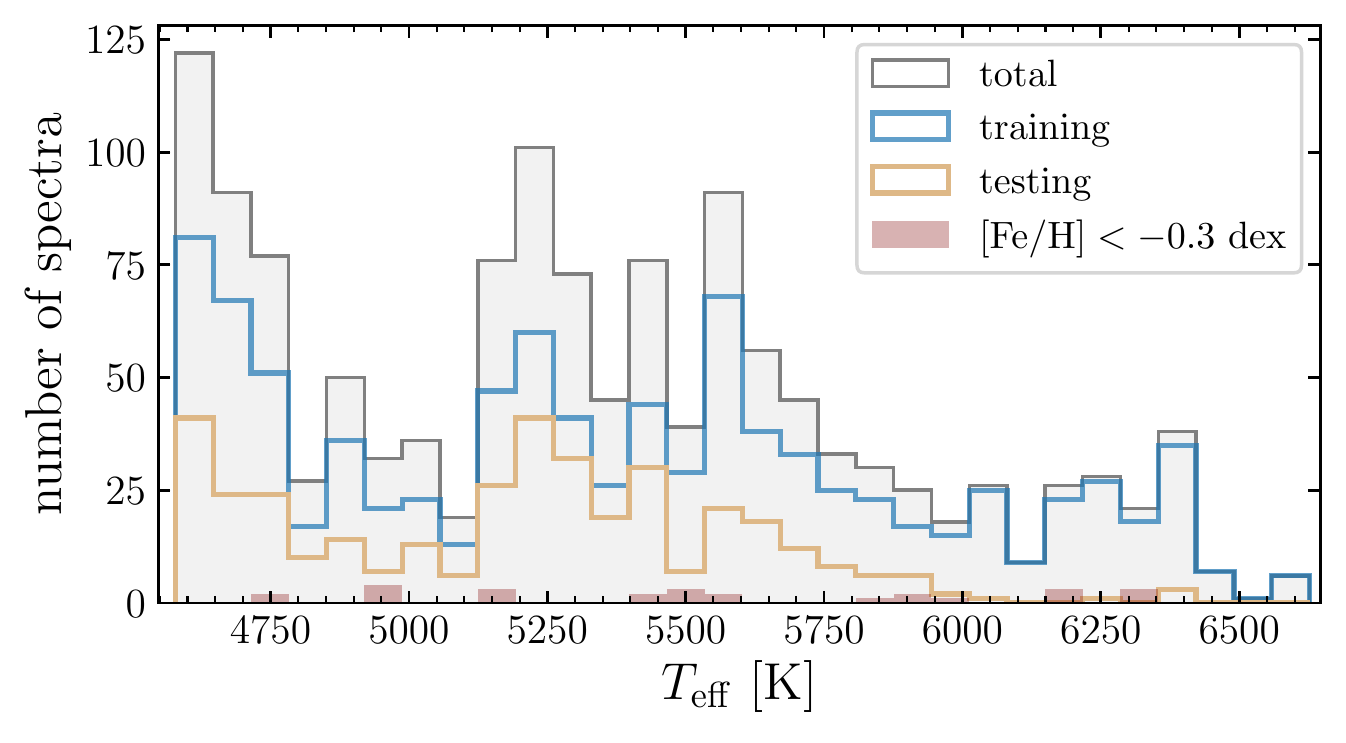}}
      \caption{Temperature distribution of the spectra in our dataset, shown as a gray histogram. The temperature distributions of the training and testing samples are shown in blue and pale brown, respectively. We also indicated in red the temperature distribution of the spectra of stars with |Fe/H$]<-0.3$~dex that did not enter the training.}
      \label{Fig:Teff_distribution}
\end{figure}
\begin{figure*}[!ht]
    \centering
    \resizebox{\hsize}{!}{\includegraphics{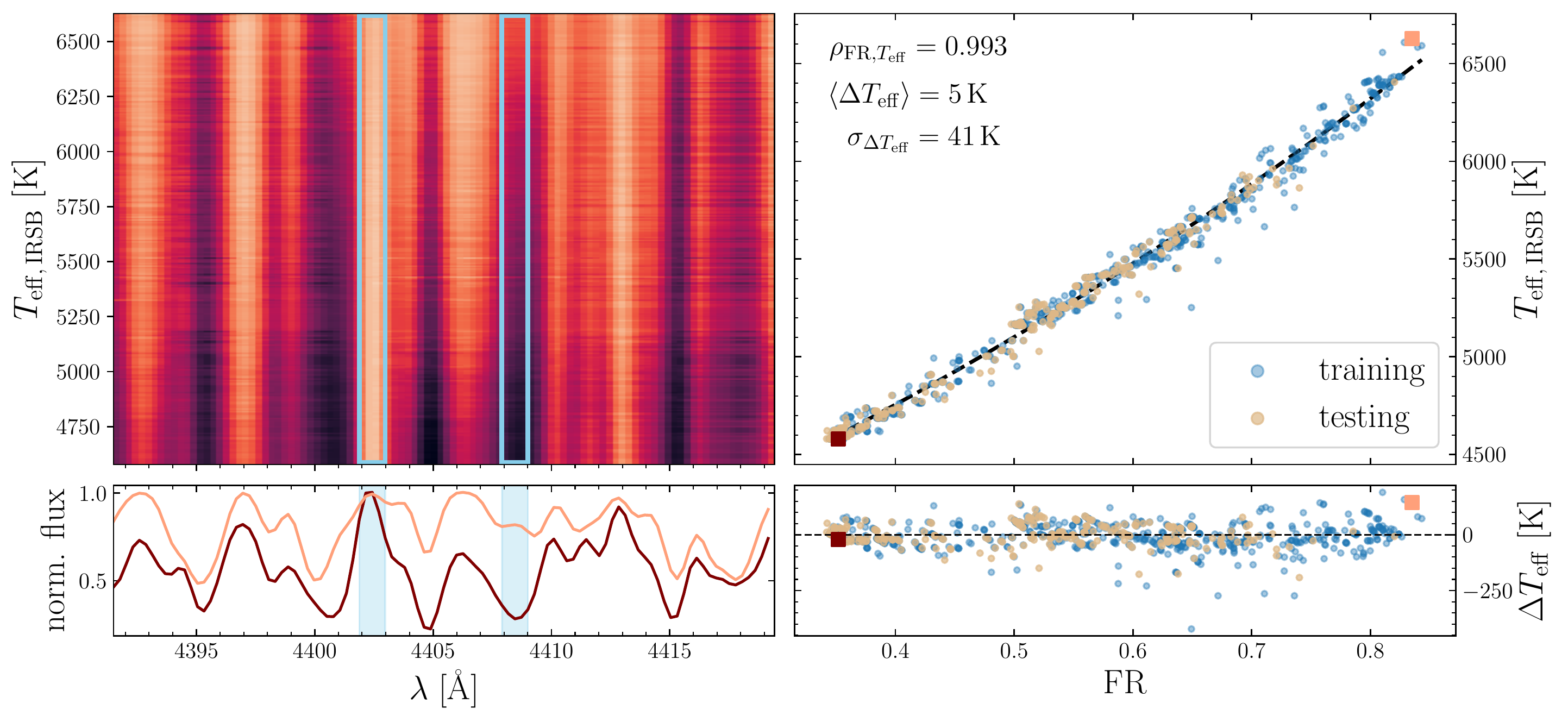}}
      \caption{Properties of an individual relation using flux ratios around 4405~\AA. The \textit{top left panel} shows the evolution of the neighboring spectral domain with \Teff, where darker colors indicate deeper features. In the \textit{bottom left panel}, the coolest (dark brown) and hottest (orange) spectra at our disposal are displayed. For illustration purposes, the spectra were pseudo-normalized by the 99$^\mathrm{th}$ percentile of the flux in the presented range. In both panels, the regions considered for computing the flux ratios are marked in clear blue. The \textit{top right panel} shows the analytical relation (dashed black curve) together with the \Teff{} labels for the training (blue) and testing (pale brown) samples. Using the same color coding, the \textit{bottom right panel} shows the difference between the temperatures computed using the individual analytical relation and the input labels. In both panels, the coolest and hottest spectra are again depicted in dark brown and orange, respectively. 
      }
      \label{Fig:relation_demo}
\end{figure*}

\par For fitting, the dataset was split into a training set (70\% of the sample) and a holdout (or testing) dataset (30\%), which was used for unbiased estimates of the model fitness. In principle, it would be desirable to have at the same time both a homogeneous coverage in label space and the training and testing sets obeying identical probability distributions. In practice, however, since our \Teff{} distribution is skewed towards lower temperatures (see Fig. \ref{Fig:Teff_distribution}), we decided to foster \Teff{}-homogeneity for the training sample by randomly drawing with the inverse temperature density as weight. As a consequence, the holdout set follows a slightly different probability distribution, which is indicated in Fig. \ref{Fig:Teff_distribution}. The only alternative approach would be to disregard a substantial number of spectra in the intermediate- and low-temperature regime from the analysis to establish equal numbers throughout the whole \Teff{} range. Yet, we favor our choice for two reasons: First, some spectra have a discontinuous wavelength coverage due to chip gaps and others are limited to the visual range and do not extend to the near-infrared CaT region (8498--8662~\AA). Hence, not all FRs can be measured in all spectra, which adds an additional layer of complication that is best mitigated by having a higher number of spectra from different spectrographs. The second reason for keeping all observed spectra is that it enables a better coverage in fundamental parameters other than \Teff{}, such as [Fe/H], $\log{g}$, or period, which are key to spotting degeneracies with these higher-order parameters.\\

\par Even though we used correlation coefficients for feature selection and therefore technically pre-filtered our features for close-to linear relations, we allowed our model to be more flexible by adding the possibility of curvature through second-order polynomials. It is noteworthy at this point that having such a simple model renders a validation set for tuning hyper-parameters superfluous. One such case is presented in Fig. \ref{Fig:relation_demo}, where we, on the one hand, exemplarily demonstrate how our most well-constrained spectral region (see next section for the selection criteria) around $\sim4405$~{\AA} behaves with temperature, and, on the other hand, we illustrate how the identified FR constrains \Teff{} both in the training and testing sets.\\

\begin{figure}[!h]
    \centering
    \resizebox{\hsize}{!}{\includegraphics{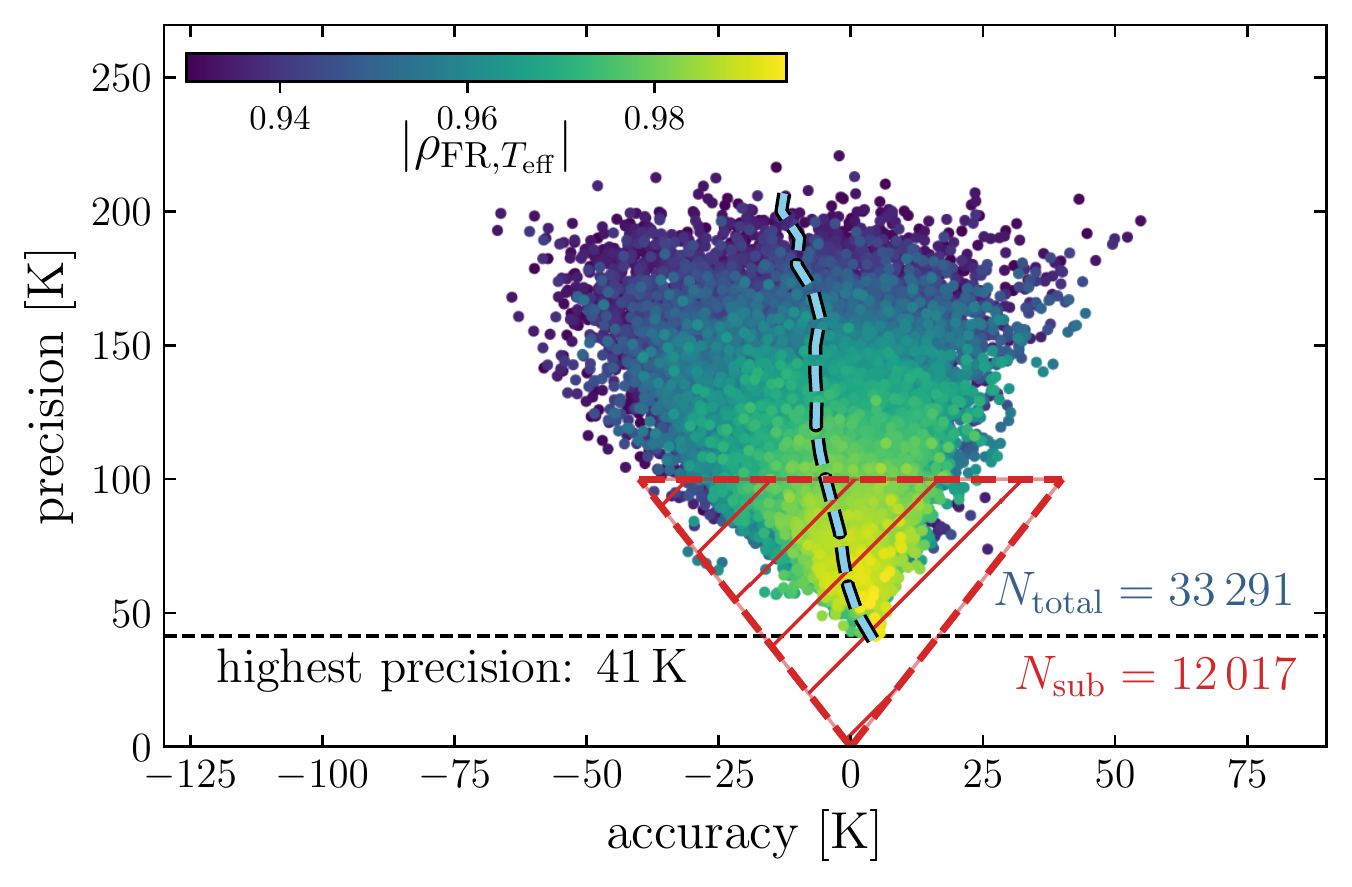}}
      \caption{Precision and accuracy (in K) for all the relations retrieved by the analysis, color-coded by their correlation coefficient. Only relations for which the precision is better than 100~K and less than 3 $\times$ the accuracy are further considered (red triangle). The dashed line shows a slight accuracy bias (see text) as a function of precision, which was computed using a boxcar median filter.}
      \label{Fig:pre_selection}
\end{figure}
\par The actual training was performed on each of the 33\,291 FRs that survived the feature selection. While the model was fit using the training set, its fitness in terms of accuracy (mean deviation) and precision (standard deviation, $\sigma$) was independently determined from the testing set. To lessen the impact of artefacts on individual FRs induced by, for example, erroneous phasing, wavelength calibration errors, and cosmics, we allowed a clipping of up to five of the most extremely deviating spectra both from the training and testing sets. The distribution of the respective findings for the individual features is depicted in Fig. \ref{Fig:pre_selection}. 

\par We found that there is a slight accuracy bias that tends towards negative values with decreasing precision. This could be owed to the circumstance that for any set of two physically identical spectra in the training and testing sets, respectively, the testing spectrum shows an offset \Teff{} label. Such an effect could occur if these spectra are associated with different Cepheids, where inter-target systematics (e.g., reddening, projection factor, etc.) could hamper the labeling in the IRSB analysis. Nevertheless, given that the systematic bias remains well below 25~K, we do not investigate this behavior further (but see Sects.~\ref{res_teff} and \ref{dis:red}). 

\begin{figure*}[!h]
    \centering
    \resizebox{\hsize}{!}{\includegraphics{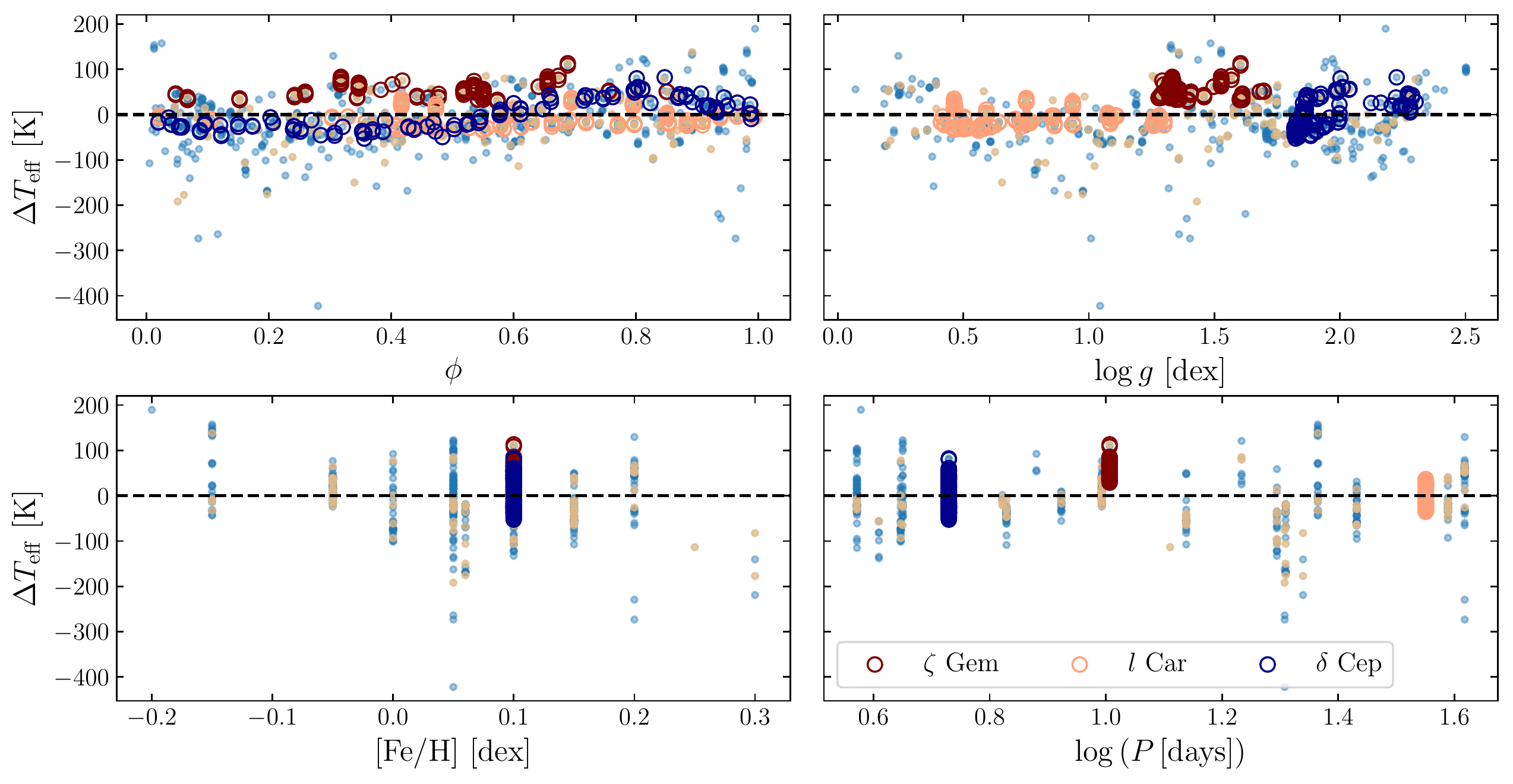}}
      \caption{For the same relation as in Fig. \ref{Fig:relation_demo}, the residuals $T_\mathrm{eff,\,IRSB}-T_\mathrm{eff,\,FR}$ are shown as a function of phase ($\phi$, \textit{top left}), \logg{} (\textit{top right}), [Fe/H] (\textit{bottom left}), and $\log{P}$ (\textit{bottom right}). Colored open circles highlight spectra for the targets $\zeta$~Gem, $l$~Car, and $\delta$~Cep.}
      \label{Fig:relation_demo_residuals}
\end{figure*}
\par Not surprisingly, the precision correlates with the correlation coefficient, although -- due to allowing for curvature in the model -- we deem the precision to be the more desirable quantifier for the goodness of the model fit.

\subsection{Final selection of flux ratios}
\label{final selection}
\par After having fit all candidates for useful features, we imposed further quality cuts, in a sense that only such features were kept, for which the model assessment yielded $\mathrm{precision/accuracy}>3$ and the precision remained below 100~K. These criteria are superimposed in Fig. \ref{Fig:pre_selection} and resulted in 12\,017 FRs for further analyses.

\par Next, we had to make sure that the final selection of features is linearly independent. Only then, the predicted labels from all models and unseen data can be averaged in a meaningful way, that is, without having to deal with inevitable (unknown) covariances\footnote{We note, in particular, that some lines are used simultaneously in several LDR relations \protect\citep{Proxauf2018}, their depth being employed sometimes in the numerator and sometimes in the denominator of the ratio. For a detailed comparison of the IRSB labels used in this paper and the LDR method, we refer the reader to paper II in this series.}. We achieved this by ranking the features by their associated model precision and subsequently exclude those that have pixel overlaps with a feature of higher precision. This procedure further reduced the number of leftover features to 247, which indicates that the vast majority are clustered around a limited number of \Teff{}-sensitive line profiles and are only shifted by incremental $\lambda$ steps with respect to each other.\\

\par Finally, the remaining 247 relations were visually checked for erratic behaviors in the residuals $T_\mathrm{eff,\,IRSB}-T_\mathrm{eff,\,FR}$ to exclude (strong) influences on the FRs from parameters in excess of \Teff{}. Another considered factor was substructure that is indicative of an insufficient description by our quadratic model (e.g., discontinuities or non-flat residuals as a function of FR). For our best relation, the residuals are presented with respect to $\phi$, $\log{g}$, [Fe/H], and $\log{P}$ in Fig. \ref{Fig:relation_demo_residuals}. There, we also highlight the Cepheids $\zeta$~Gem, $l$~Car, and $\delta$~Cep as representatives with good phase and $\log{g}$ coverage as well as very different pulsation periods. It is evident that there is a persistent, statistically significant bias of $\sim59$~K between $\zeta$~Gem and $l$~Car ($\langle\Delta T_\mathrm{eff}\rangle=56\pm16$~K and $-3\pm19$~K, respectively, both only computed from the testing set) irrespective of phase and $\log{g}$. We would like to stress at this point that this shows that what we call precision of a relation (i.e., the scatter measured from test spectra of many different Cepheids) is largely governed by star-to-star systematics in the labels and therefore encompasses external errors. In fact, the internal precision of a single of our FR relations as deduced from one target alone (in the example above $\zeta$~Gem and $l$~Car) resides below 20~K. The reader is referred to Sect. \ref{res_teff} where we discuss in detail the implications of labeling systematics on the full sample of relations. After our visual sanity check, we are left with 143 relations.

\section{Results}
\label{results}

\par After carefully selecting and phasing the dataset, we trained our relations using IRSB \Teff{} labels and obtained 143 calibration relations that can be used to determine the effective temperature of Cepheids. Before describing the properties of the identified relations, we want to test the stability of individual relations with respect to the signal-to-noise ratio and continuum variations, and briefly discuss the remaining systematics on the final temperature.

\subsection{Signal-to-noise ratio stability}
\label{Sect: SNR stab}

\begin{figure}
    \centering
    \resizebox{\hsize}{!}{\includegraphics{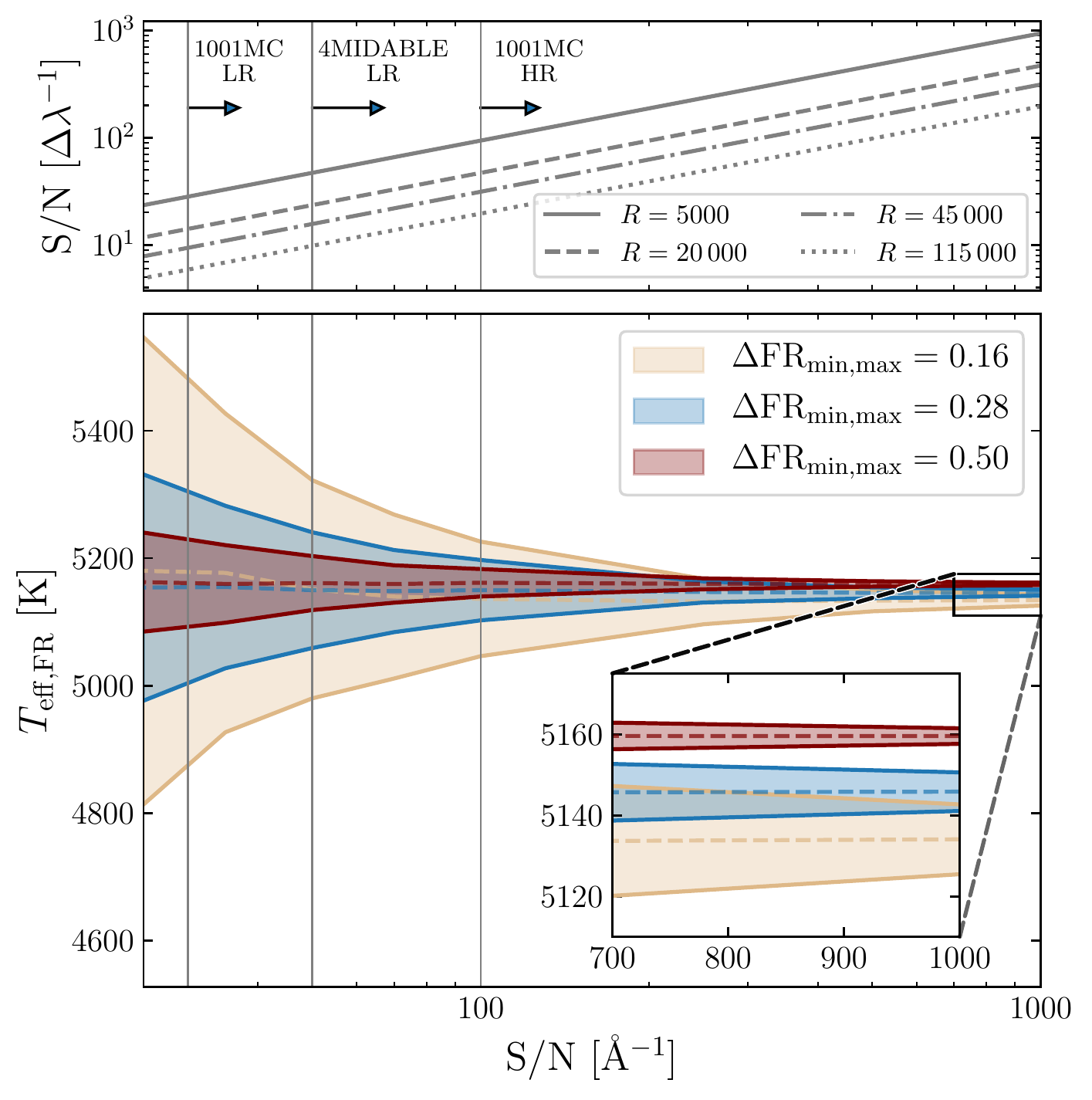}}
      \caption[]{Stability against S/N per {\AA} for three relations of different maximum FR excursion (cf., color coding and legend in the \textit{lower panel}). The corresponding S/N per resolution element for four different resolving powers (see legend) are given in the \textit{top panel}. Shown are the measured mean values (colored dashed lines) and $1\,\sigma$ standard deviations (solid lines enclosing filled areas) from an MC simulation with artificially injected noise (see main text for details). The zoom-in view emphasizes the (marginal) relation-to-relation inaccuracies (see middle panel of Fig. \ref{Fig:FR_distribution}).
      Vertical lines indicate the minimum S/N requirements for different 4MOST consortium surveys: 
      $\mathrm{S/N}=50~\mathrm{\AA}^{-1}$ in the magnesium or calcium triplet regions
      (4MIDABLE-LR\protect\footnotemark);
      $\mathrm{S/N}=30~\mathrm{\AA}^{-1}$ in the magnesium or calcium triplet regions (1001MC\protect\footnotemark, LR);
      $\mathrm{S/N}=100~\mathrm{\AA}^{-1}$ at  6200~\AA{} (1001MC, HR;
      4MIDABLE-HR\protect\footnotemark).
      }
      \label{Fig:noise_stability}
      \protect
\end{figure}
\addtocounter{footnote}{-3}
\stepcounter{footnote}\footnotetext{4MIDABLE-LR: \protect\cite{Chiappini2019}}
\stepcounter{footnote}\footnotetext{1001MC: \protect\cite{Cioni2019}}
\stepcounter{footnote}\footnotetext{4MIDABLE-HR: \protect\cite{Bensby2019}}

\par Given a high-S/N (e.g., $>100~\mathrm{\AA}^{-1}$) spectrum, even a single of our discovered features is capable of predicting \Teff{} at a precision that is more than sufficient for subsequent abundance analyses (i.e., $\sigma_{T_\mathrm{eff,\,FR}}<<100$~K). In practice, for larger samples this is only feasible for nearby and therefore bright Cepheids for which spectra can be obtained within reasonable exposure times and/or 8+\,m-class telescopes. In light of the fact that upcoming spectroscopic campaigns such as WEAVE and 4MOST at 4\,m-class facilities will only allow for a more restrictive spectrum quality ($\mathrm{S/N}\gtrsim30~\mathrm{\AA}^{-1}$), our extended number of relations with a wide wavelength coverage is paramount. In order to optimize the averaging strategy for predicted labels from multiple features, knowledge about the S/N stability is desirable.\\

\par The parameter affecting the noise sensitivity most is $\Delta\mathrm{FR}_\mathrm{min,\,max}$, which denotes the excursion an FR covers over the full range of \Teff{}. To test this, we have conducted a Monte Carlo (MC) analysis using a single-epoch spectrum of $\zeta$~Gem with attributed parameters ($\phi=0.48$, $T_\mathrm{eff}=5206$~K, $\log{g}=1.29$~dex, [Fe/H$]=+0.10$~dex) that are roughly consistent with the midpoint of the parameter coverage of our dataset (cf., Fig.~\ref{Fig:sample_HRD}). The spectrum was obtained with $R\approx115\,000$ and $\mathrm{S/N}=276$~pixel$^{-1}$ using HARPS. At 0.01~{\AA} per pixel, this latter value corresponds to $\mathrm{S/N}\approx2760$~{\AA}$^{-1}$, which leads us to the conclusion that for tests with artificial noise satisfying $\mathrm{S/N}\leq1000$~{\AA}$^{-1}$ contributions from the real observed noise level can be safely neglected. To estimate both mean values and standard deviations at various S/N values between 25 and 1000~{\AA}$^{-1}$, per S/N realization, we generated 500 spectra with injected random noise, degraded each to $R=5000$, measured FRs, and predicted the \Teff{} labels. 

\par The result can be seen in Fig. \ref{Fig:noise_stability} where we indicate the course of $\sigma_{T_\mathrm{eff,\,FR}}$ 
as a function of S/N for three representative features with different realizations of $\Delta\mathrm{FR}_\mathrm{min,\,max}$; the lowest allowed spacing of 0.15, an intermediate value (0.28), and a large excursion (0.50). For these three relations, the noise-induced standard deviation ranges from (366~K, 177~K, 77~K) at $\mathrm{S/N}=25~\mathrm{\AA}^{-1}$ through (90~K, 47~K, 22~K) at $\mathrm{S/N}=100~\mathrm{\AA}^{-1}$ to (9~K, 5~K, 2~K) at $\mathrm{S/N}=1000~\mathrm{\AA}^{-1}$. Hence, even at $\mathrm{S/N}=25~\mathrm{\AA}^{-1}$, from a mathematical point of view -- i.e., neglecting artefacts that real spectra inevitably suffer from -- averaging only 16 of our relations with the strongest (worst) S/N sensitivity already yields internal temperature precisions that are sufficient for meaningful abundance analyses ($366~\mathrm{K}/\sqrt{16}=92$~K). 

\par An important -- though not unexpected -- observation is that, considering the scatter, the mean $T_\mathrm{eff,\,FR}$ is perfectly stable with S/N. This highlights the superiority of our method over many already existing analysis tools that show systematically increasing over- or underestimations of the continuum level with varying S/N. As a consequence, further down the line such tools introduce potentially strong systematic scatter when comparing results (e.g., [Fe/H] or other chemical abundances) obtained from spectra of substantially varying quality (see for instance Fig.~2 in \citealt{Reichert2020}.

\subsection{Stability against continuum variations}
\label{cont var}
\begin{figure}
    \centering
    \resizebox{\hsize}{!}{\includegraphics{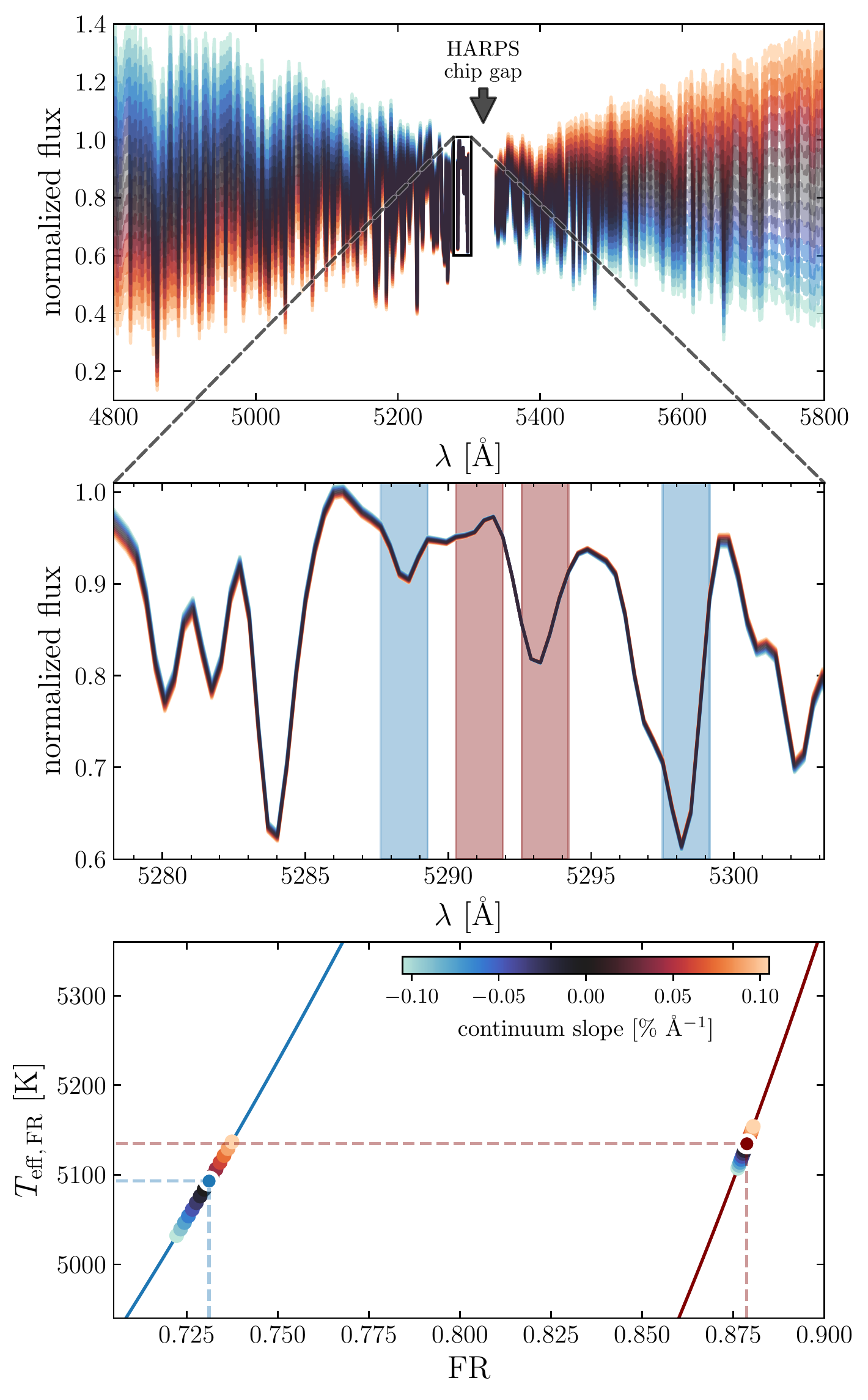}}
      \caption{Impact of continuum variations on FRs from close- (red) and wide-spaced (blue) wavelength components. \textit{Upper panel}: 15 times the same $\zeta$~Gem spectrum as was used in Sect. \ref{Sect: SNR stab} with varying continuum slopes (see color bar in \textit{lower panel}). The missing part in the presented spectrum corresponds to HARPS' chip gap. \textit{Middle panel}: zoomed-in view of the region around $\sim5295$~{\AA} where the $\lambda$ regions are highlighted by vertical spans. \textit{Lower panel}: deduced $T_\mathrm{eff,\,FR}$ values as a function of the measured FRs at different continuum slopes. The models are represented by red and blue solid curves while dashed red and blue lines and filled circles of the same colors indicate the finding from the original spectrum without any modification of the continuum.  
              }
      \label{Fig:continuum_stability}
\end{figure}
Another investigated source of error is the stability of \Teff{} predictions from our models against extreme variations of the continuum level between two involved wavelength ranges due to, for instance, a mistreated removal of the instrument's blaze function. As opposed to noise-induced errors -- which can be reduced by averaging a number of relations -- potential errors from this are of systematic origin and ultimately affect our accuracy. In Fig. \ref{Fig:continuum_stability} we illustrate how the effect differs for two FRs at the extreme ends of the distribution of wavelength spacings, $\Delta \lambda$. We employed the same spectrum of $\zeta$~Gem introduced in the previous section and multiplied linear continua with slopes as drastic as $\pm0.1$\,\%\,{\AA}$^{-1}$. Not even the steepest slopes exceed a $\pm25$~K effect for the closely-spaced case, whereas a maximum excursion of $\pm50$~K was found for the largest $\Delta \lambda$. One might desire to drop or downweight the ratios with largest $\Delta \lambda$ when analyzing high-dispersion spectra with narrow orders, since they are potentially affected by somewhat stronger continuum variations than other devices.

\subsection{Size and origin of remaining systematics}
\label{res_teff}
\begin{figure*}
    \centering
    \resizebox{\hsize}{!}{\includegraphics{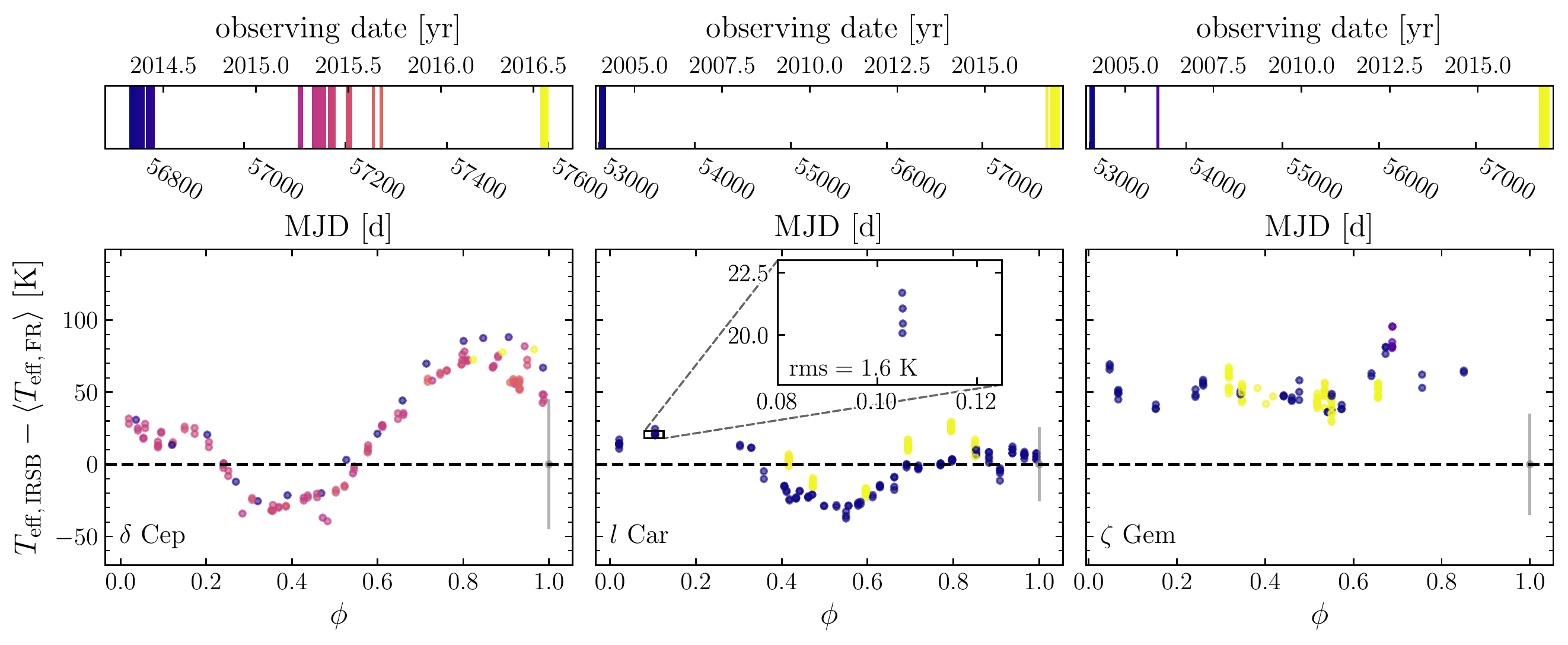}}
      \caption{Representative examples of the three distinct residual patterns as discussed in the text: $\delta$~Cep (IRSB issues seemingly constant with MJD, \textit{left panel}); $l$~Car (IRSB issues with MJD-dependence, \textit{middle panel}); $\zeta$~Gem (constant offset possibly due to erroneous reddening, \textit{right panel}). Shown are the median residuals of the best 100 trained relations that were measured in individual spectra with respect to the input labels. The abscissa indicates the pulsation phase, $\phi$. As in Fig. \ref{Fig:rephasing_demo_RSPup}, the color coding resembles the time of observation of the individual spectra (cf., \textit{upper panels}). Typical values for the $1\,\sigma$ scatter are depicted by gray error bars. The inlay in the \textit{middle panel} highlights five spectra of $l$~Car that were taken with HARPS back to back within less than 30~min. The rms scatter of 1.6~K underlines the internal precision of our method.}
      \label{Fig:traning_vs_labels_3cases}
\end{figure*}

\par In the two previous sections, we investigated individual relations. From now on, we are dealing with the average \Teff{} derived from a large number of relations, and not anymore from a temperature derived from a single relation.

\par Thanks to our efforts to minimize systematics, we are left with small-scale residuals (a few tens K) that show different patterns, displayed in Fig.~\ref{Fig:traning_vs_labels_3cases}. The first pattern is a phase-dependent offset that seems independent from the MJD of the observation, at least for the relatively small range of MJDs considered here (case of, e.g., $\delta$~Cep). The residuals are likely related to the IRSB method. We checked, however, that there is no phase shift between the optical and near-infrared photometry, nor with the radial velocity data \citep[see][and references therein]{Storm2011a,Storm2011b}. Since the peak of the deviation happens at maximum and minimum radii, we also investigated whether the radius amplitude could be influenced by the value selected for the $p$-factor. Using $p$=1.3 instead of the nominal value ($p$=1.41), we found that the effect on $T_\mathrm{eff,\,IRSB}$ is less than 2~K 
over the entire phase range, thus excluding the $p$-factor from the possible source for the residuals. The second pattern is a similar phase-dependent offset, this time MJD-dependent (case of, e.g., $l$~Car). Although the residuals are again likely related to the IRSB analysis, their variation probably originates from an inaccurate rephasing or from cycle-to-cycle variation of the temperature curve. The third pattern resembles a constant offset between the retrieved temperature, $T_\mathrm{eff,\,FR}$, and the labels (case of, e.g., $\zeta$~Gem), possibly due to an erroneous value for a global, phase-independent parameter (reddening?) in the IRSB analysis (see Sect.~\ref{IRSB}).

\subsection{Properties of the identified relations}
\label{res_rel}

\par
\begin{figure}[!t]
    \centering
    \resizebox{\hsize}{!}{\includegraphics{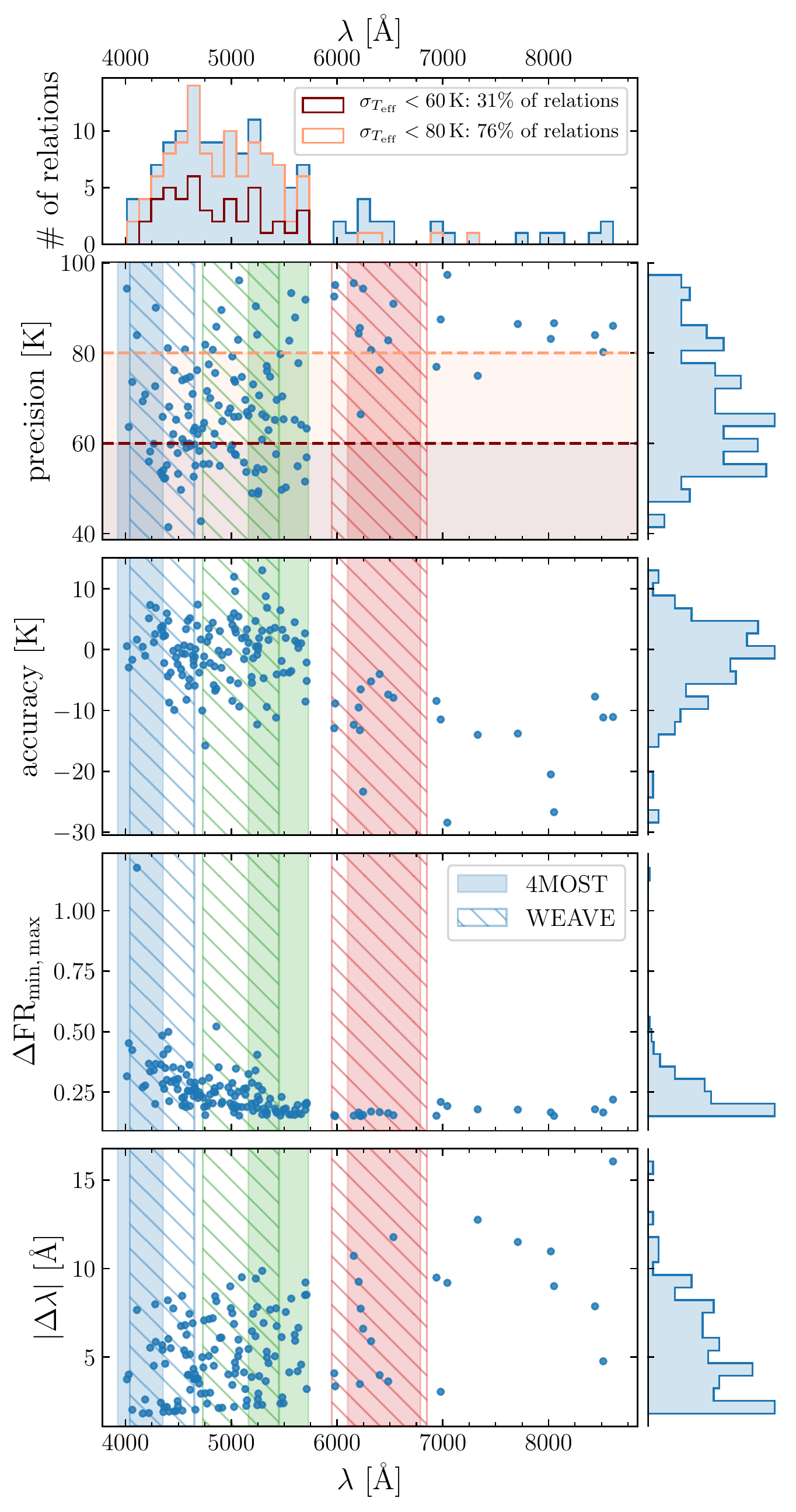}}
      \caption{Distribution of the properties of FRs as a function of wavelength. Blue, green, and red vertical spans depict the three spectral windows of 4MOST (filled) and WEAVE (hatched) high-resolution spectrographs. \textit{Upper panel}: Number of independent relations to derive \Teff{} from flux ratios per wavelength range. \textit{2$^{nd}$ panel}: Precision of individual relations to derive \Teff{} from flux ratios as a function of wavelength. \textit{3$^{rd}$ panel}: Accuracy of individual relations to derive \Teff{} from flux ratios as a function of wavelength. 
      \textit{4$^{th}$ panel}: Flux ratio excursion, $\Delta\mathrm{FR}_\mathrm{min,\,max}$, as a function of wavelength.
      \textit{Bottom panel}: Wavelength spacing $\Delta \lambda$ of the two components of a ratio as a function of wavelength.}
      \label{Fig:FR_distribution}
\end{figure}

 It is not a surprise that the majority of the flux ratios involves features in the blue region of the spectra (Fig.~\ref{Fig:FR_distribution}, top panel), given the wealth of potentially temperature-sensitive metallic lines in this spectral domain. However there is still a good number of relations in the [5000--6000]~\AA{} range, and more specifically around the magnesium triplet (MgT) and the H$\alpha$ line. A handful of ratios will certainly prove very useful in the CaT region. 

There are only three relations overlapping the spectral range of the Gaia Radial Velocity Spectrometer \citep[RVS,][]{Sartoretti2018}. As shown in the second and third panels of Fig.~\ref{Fig:FR_distribution}, their precision and accuracy are among the lowest within our set of relations, probably because only a small fraction of the spectra in our sample cover this wavelength range. Such a paucity would also prevent a proper outlier rejection in case of undesirable artifacts in the analyzed spectra. At this stage, we would therefore caution against using these three lines solely, for instance in order to analyze Gaia RVS spectra. However, training at the resolution of the RVS ($R=11\,500$) might provide more calibration relations with better accuracy and precision. This will be investigated in a forthcoming paper.

\par
\begin{figure}
    \centering
    \resizebox{\hsize}{!}{\includegraphics{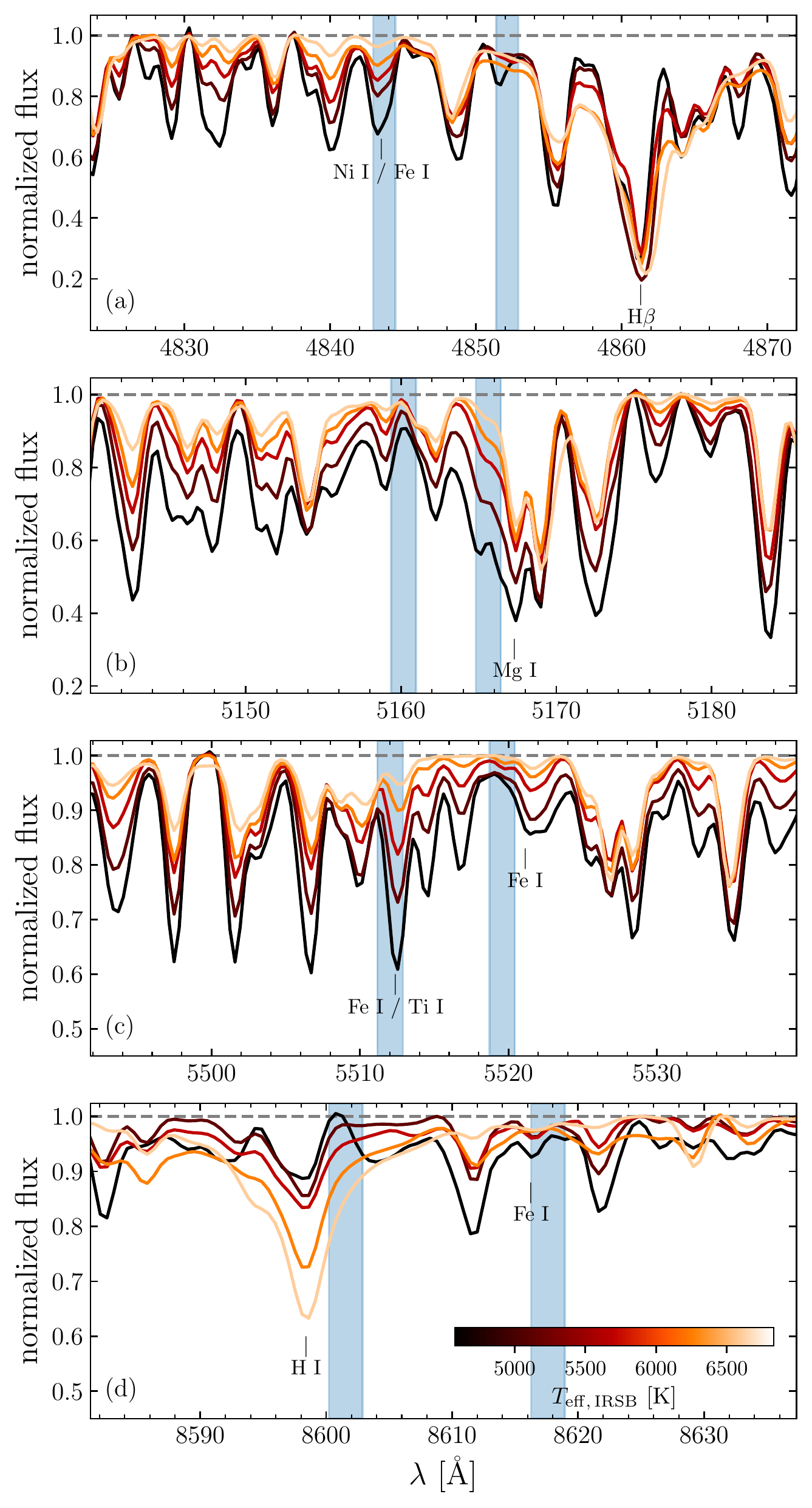}}
      \caption{Illustration of a few spectral domains where \Teff-sensitive flux ratios have been recovered: a) blue wing of H$\beta$, b) MgT region, c) metallic lines around 5510~\AA, d) \ion{H}{i} line in the near-infrared. See Sect.~\ref{res_rel} for details. Spectra are color-coded by temperature, where darker colors indicate cooler temperatures.
      For illustration purposes, the spectra were pseudo-normalized by the 99$^{th}$ percentile of the flux in the presented range.}
      \label{Fig:four_lambda_regions}
\end{figure}

 In Fig.~\ref{Fig:four_lambda_regions}, we display a few interesting spectral regions harboring flux ratios:
\begin{enumerate}[nosep, label=(\alph*)]
    \item The first feature we show is located in the blue spectral region, in the vicinity of H$\beta$. FRs involving the wings of hydrogen lines are typical \Teff{} indicators as already found by \citetalias{Hanke2018}. A \Teff-sensitive feature (the blend of a \ion{Ni}{i} and a \ion{Fe}{i} line) is normalized by another spectral domain in the blue wing of H$\beta$, which varies much less with \Teff.
    \item The second example is located in the MgT region (5167-5184~\AA). One of the spectral domains is located directly in the bluemost of the \ion{Mg}{i} lines of the Mgb triplet and varies notably with \Teff{} while the other one is only slightly temperature-sensitive.
    \item Another example involves metallic lines, where the blend of a \ion{Fe}{i} line and a \ion{Ti}{i} line shows a strong dependence on \Teff{}, which seems to be compensated by the wing of a \ion{Fe}{i} line whose sensitivity on \Teff{} is much less pronounced and might also correct for the effect of another parameter like \logg{} or [Fe/H].
    \item Finally, we show a flux ratio with two components located close to a \ion{H}{i} line in the near-infrared. It does not have a typical behavior, as both components have an opposite sensitivity to \Teff: the first one is located in the wing of the \ion{H}{i} line, which becomes stronger when \Teff{} increases, while the other one lies in the wing of a \ion{Fe}{i} line, which becomes more prominent when \Teff{} decreases.\\
\end{enumerate}

\par Before examining the second panel of Fig.~\ref{Fig:FR_distribution}, we wish to recall the argument already made at the end of Sect.~\ref{final selection}: what we call precision is the scatter measured from test spectra of many different Cepheids. It combines two effects, of which star-to-star systematics is the dominant one, while the internal precision of an individual relation as deduced from a single target falls below 20~K. Keeping this in mind, we see that the precision on individual relations, capped at 100~K by construction, is actually much better (Fig.~\ref{Fig:FR_distribution}, second panel). Indeed roughly half of the relations have an individual precision better than 80~K, and the best relations have precisions ranging from 40 to 60~K; they cover, however, a more restricted wavelength domain.
\par When combining all relations, we are able to trace \Teff{} variations of a few Kelvins. For instance, the middle bottom panel of Fig.~\ref{Fig:traning_vs_labels_3cases} indicates the outcome of our analysis when investigating consecutive short exposure spectra of $l$ Car (P$\sim$35.5d) taken with HARPS within less than 30~min. We achieved a rms scatter of 1.6~K (using only the 100 best relations), which underlines the internal precision of our method.

\par Although our criterion was slightly looser ($\pm$40~K), the accuracy of our relations mostly falls within $\pm$10~K (Fig.~\ref{Fig:FR_distribution}, third panel) and rarely exceeds 20~K, thus ensuring that we reproduce almost exactly the \Teff{} scale of the input labels. From a practical point of view, we would recommend to make use of only a selected number of the best relations (since their individual accuracy and precision are known) rather than sigma-clipping temperatures derived from individual relations, if the goal of the operation is to improve the precision of the final \Teff. Note that this will probably come at the cost of the wavelength coverage, as shown in the second and third panel of Fig.~\ref{Fig:FR_distribution}. An alternative (possibly better) approach would be to perform a weighted median using the (known) inverse variance (i.e., squared precision) of individual relation as weight.

\par The last two panels of Fig.~\ref{Fig:FR_distribution} display the excursion $\Delta\mathrm{FR}_\mathrm{min,\,max}$ covered by flux ratios, and the wavelength spacing $\Delta \lambda$ of the two components of the flux ratios, respectively. They both can be interpreted by the larger occurrence of temperature-sensitive features in the blue regions of Cepheids' spectra.

\section{Discussion}
\label{discussion}

\subsection{Applicability at low metallicity}
\label{low_met}

\begin{figure*}[!ht]
    \centering
    \resizebox{0.9\hsize}{!}{\includegraphics{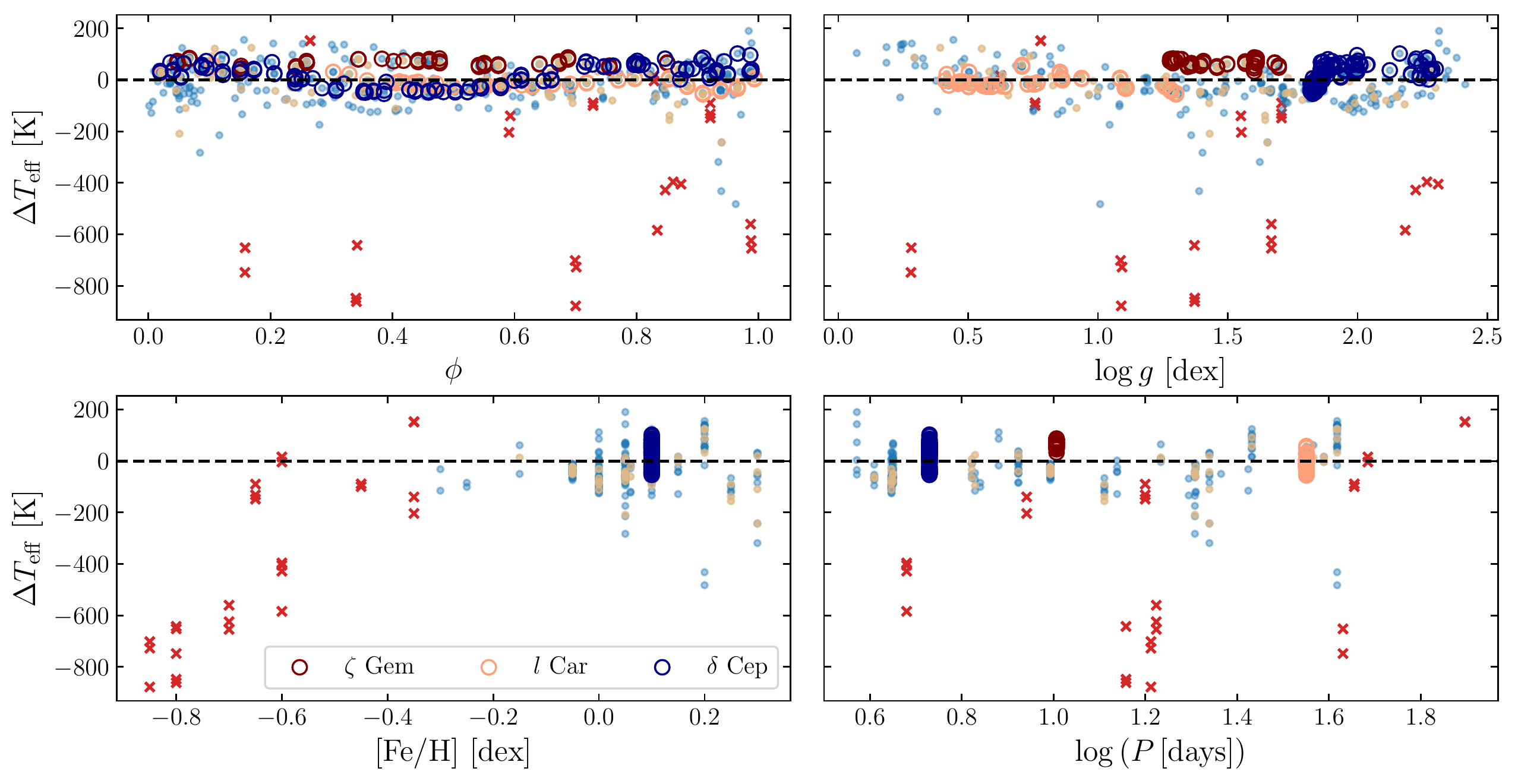}}
      \caption{Same as Fig.~\ref{Fig:relation_demo_residuals}, where \Teff{} derived for low-metallicity SMC Cepheids are shown by red crosses.}
      \label{Fig:placeholder}
\end{figure*}

\par We used the small number of spectra of SMC Cepheids at our disposal to test the applicability of our method at low metallicity. Since the spectra do not cover the entire wavelength domain, not all the relations could be investigated. For those that could be tested, we see, however, a clear pattern exemplified in Fig.~\ref{Fig:placeholder}: the relations hold down to [Fe/H]$\sim$-0.6~dex and the residuals quickly increase below this value, reaching $\approx$800~K at [Fe/H]$\sim$-0.8~dex.

\par This is not a surprise, as the training sample does not contain any low-metallicity SMC Cepheids (see Sect.~\ref{train}). In its current stage, the method is then applicable to Galactic Cepheids in the outer disk and to a large fraction of the LMC Cepheids, but not to the bulk of the SMC ones. Here we want to stress that the current values of [Fe/H] have been obtained using a \Teff{} scale based on line depth ratios, and could therefore be modified when adopting temperatures derived from our flux ratios.

\par Finally, we note that the lack of SMC Cepheid spectra not only deprives us from a training sample at low metallicity, but also decreases the quality of the spectral phasing: with only a few consecutive spectra at hands, we cannot investigate possible period changes or cycle-to-cycle variations, which are to be expected since the current sample is biased towards brighter, longer-period Cepheids.

\par Extending our \Teff{} scale to low-metallicity Cepheids would require a large number of spectra of SMC Cepheids. Including such spectra in the training will certainly lead to a different (presumably smaller) set of (more) universal relations. A larger number of relations may however be recovered by using higher-order/more-dimensional models.

\subsection{A new (spectroscopic) method for determining the reddening of Cepheids ?}
\label{dis:red}
\begin{figure}[!ht]
    \centering
    \resizebox{0.9\hsize}{!}{\includegraphics{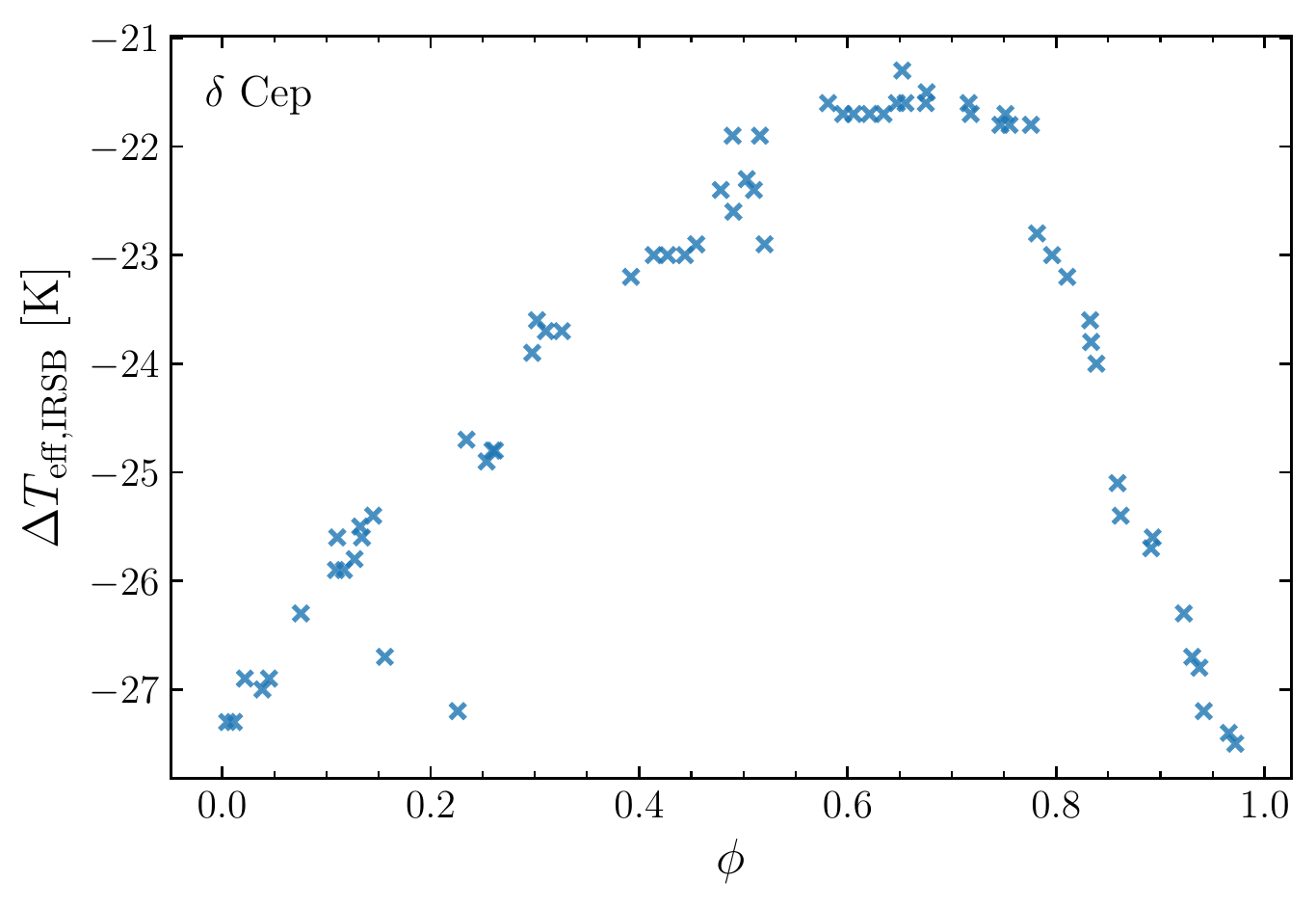}}
      \caption{Phase-dependent shift (modified $-$ adopted) of the IRSB temperature of $\delta$~Cep when computed either with the modified $E(B-V) =0.069$~mag or with $E(B-V)=0.075$~mag as adopted throughout this work.}
      \label{Fig:deltaCep_modEBV}
\end{figure}
\par Different parameters can limit the efficiency of the various flavors of IRSB methods in deriving accurate distances for Cepheids, namely the limb darkening correction \citep[see][]{Sabbey1995,Neilson2013}, the possible dependence of the projection factor on period (see \citealt{Groenewegen2013} and \citealt{Kervella2017} for instance) or metallicity \citep[discarded by e.g.,][]{Nardetto2011}, the presence of a circumstellar envelope \citep[e.g.,][]{Kervella2006,Hocde2020}, or the reddening.
\par If one assumes that reddening is the dominant uncertainty in the determination of \Teff{} as a by-product of IRSB methods, those Cepheids with minimal reddening (which we used in our training/testing sample) provide the absolute zero-point of the temperature scale. The other Cepheids should then align on the same scale. As a consequence, any systematic offset between the temperature derived (with unprecedented precision) using flux ratios and the temperature determined using the IRSB method could be attributed to uncertainties on the reddening. Conversely, it is possible to compute such a temperature offset and, in turn, to find out how the reddening would be modified for both \Teff{} scales to match. For the sake of the argument, we conducted such an exercise for the Cepheid $\delta$ Cep and managed to compensate for a systematic shift of 23~K between $T_\mathrm{eff,\,IRSB}$ and $T_\mathrm{eff,\,FR}$ by lowering $E(B-V)$ by $-$0.006~mag (0.069 instead of the initial 0.075~mag). However, modifying $E(B-V)$ does not result in a simple, global shift of $T_\mathrm{eff,\,IRSB}$: Fig.~\ref{Fig:deltaCep_modEBV} shows the outcome of our $\delta$~Cep test case. It indicates that the difference in the IRSB temperatures computed either with $E(B-V)=0.075$~mag or with $E(B-V)=0.069$~mag is phase-dependent. Incidentally, since this phase dependence is different from the one seen in Fig.~\ref{Fig:traning_vs_labels_3cases}, and since its magnitude is different by a factor of $\approx$20, it cannot smooth out the current residuals.
It is beyond the scope of this paper to investigate this behavior further. These findings may lead to a better understanding of the current scatter of IRSB-based PL relations and/or the $p$-factor controversy.

\section{Conclusions}

\par Using an innovative, data-driven method, we inferred a \Teff{} scale for classical Cepheids tying flux ratios to temperatures based on IRSB labels. For the first time in such studies, special emphasis was placed on accurately phasing the spectra when matching them to labels. 
\par The next step to consider is to extend the \Teff{} scale towards lower, SMC-like, metallicities, which could not be done here simply because of the lack of spectra to build a training/testing sample. We also want to recover other atmospheric parameters like \logg{} or [Fe/H] using a similar technique. On a more technical side, one could envision a more refined weighting scheme, involving all four quantities: precision, accuracy, $\Delta\mathrm{FR}$, $\Delta \lambda$.
\par Since it is purely spectroscopic, the method is independent from reddening to the same extent as the IRSB labels are, and could actually offer a new path to estimate reddening values for Cepheids. The derived temperatures are precise down to a few K, and we have full control of the accuracy, We estimate that the accuracy remains better than 150~K at all phases, a value which will be greatly reduced once the uncertainties on the reddening scale, hence the \Teff/$E(B-V)$ degeneracy will be lifted. The uncertainties on $E(B-V)$ obtained via the SPIPS algorithm are of the order of 0.025~mag \citep{Breitfelder2016}. Their use may then provide a different set of labels with reduced systematic uncertainties. On the other hand, since the number of Cepheids currently within reach of interferometric measurements is very limited, this would come at the expense of number statistics, and in particular period coverage. We desire not to train an inhomogeneous sample with different sources for the labels. Rescaling relations trained on a given set of labels would be the safer way to pursue, in case there is a constant or linear shift between IRSB and interferometric labels.
\par {Our method to derive the temperature of classical Cepheids} is very flexible as it works even at low S/N, and without requiring a normalization stage. Since it is always possible to down-sample high-resolution spectra to $R=5000$ (at which the method is trained), we ensure homogeneous temperatures for both low- and high-resolution spectra, a very interesting feature with respect to large spectroscopic surveys. For instance, the stellar, circumstellar, interstellar component (SCIP) survey with WEAVE, the Milky Way Disc and Bulge Low-Resolution \citep[4MIDABLE-LR,][]{Chiappini2019}, the Milky Way Disc and Bulge High-Resolution \citep[4MIDABLE-HR,][]{Bensby2019} and the One Thousand and One Magellanic Fields \citep[1001MC,][]{Cioni2019} 4MOST surveys will all observe classical Cepheids at $R=6500$ and/or $R=20\,000$.
\par Our novel method paves the way for highly accurate and precise metallicity estimates, which will allow us to investigate the possible metallicity dependence of period-luminosity relations \citep[e.g.,][]{Romaniello2008,Bono2010a,Storm2011b} and ultimately solidify our measurement of the Hubble constant H$_{0}$.

\begin{acknowledgements}
We thank the referee, Dr. Simon Borgniet, for his useful comments which helped improving the quality of this paper. This work is partially based on observations collected at the European Southern Observatory. This work is partially based on data obtained with the STELLA robotic telescopes in Tenerife, an AIP facility jointly operated by AIP and IAC. This research has made use of the services of the ESO Science Archive Facility. This work was funded by the Deutsche Forschungsgemeinschaft (DFG, German Research Foundation) -- Project-ID 138713538 -- SFB 881 (``The Milky Way System'', subprojects A05,A08).
\end{acknowledgements}

%
%
\bibliographystyle{aa} 
\bibliography{lemasle.bib}

\end{document}